\documentclass[12pt,preprint]{aastex}

\begin{document}
\title{TIMING OBSERVATIONS OF 27 PULSARS AT THE PUSHCHINO
       OBSERVATORY FROM 1978 to 2012}
\author{T. V. Shabanova\altaffilmark{}, V. D. Pugachev\altaffilmark{},
and K. A. Lapaev\altaffilmark{}}
\affil{Pushchino Radio Astronomy Observatory, Astro Space Center,
 P. N. Lebedev Physical Institute, Russian Academy of Sciences,
 142290 Pushchino, Russia}
\email{tvsh@prao.ru}

\begin{abstract}
We present results from timing observations of 27 pulsars made at
the Pushchino Observatory over 33.5 yr between 1978 July and
2012 February. We also analyze archival Jet Propulsion Laboratory
data of 10 pulsars to extend individual data span to 43.5 yr.
We detected a new phenomenon in the timing behavior of two
pulsars, B0823+26 and B1929+10, that demonstrates a rapid change
of pulsar rotation parameters such that the sign of the second
derivative $\ddot\nu$ is reversed. An analysis of the $\ddot\nu$
changes showed that this process can be considered as a modulation
process in $\ddot\nu$.
We showed that the process of rapidly changing of pulsar rotation
parameters represents a new type of rotational irregularity
that, together with three other types of rotational
irregularities (discrete glitches, slow glitches and
quasi-periodic oscillations), forms a large-scale structure of
timing noise. These effects are all the cause of the deviation of the
timing behavior of most ordinary pulsars from a simple $\nu,\,\dot\nu$
spin-down model. We found that all four types of observed rotational
irregularities have evolving nature. Irregularities in pulsar
rotation rate pass through three evolutional stages that show
that a certain type of rotational irregularity can occur only
at a certain stage of pulsar rotation evolution.
The age boundaries between different evolutionary stages are
indistinct and diffusive. This fact is because different pulsars
having similar properties evolve along different paths. The
evolutionary scenario of the occurrence of rotational
irregularities explains well many of the observed properties of
pulsar rotation.
\end{abstract}

\keywords{pulsars: general -- stars: neutron -- stars: rotation}

\section{Introduction}
According to theoretical suggestions, a regular secular decrease
in the pulsar rotation frequency can be described by
the relation $\dot\nu\propto{-{\nu}^{n}}$,
where $n$ is the braking index \citep{man77}. For the vacuum
dipole model, when the secular spin-down is due to magnetic dipole
radiation, $n=3$. The direct way of defining the braking index,
$n={\nu{\ddot\nu}}/{{\dot\nu}^{2}}$, is based on a measurement of
the second derivative $\ddot\nu$ from the pulsar observations.
The early timing observations showed that the rotation rate
of many pulsars is subject to irregularities such as glitches
(discrete jumps in the rotation rate)
and timing noise (variations in the pulse arrival times).
These irregularities disturb the steady secular spin down and
hinder a measurement of the deterministic values of $\ddot\nu$.
For ordinary pulsars, the values of $\ddot\nu$ due to slowdown
are very small in comparison to the measurement uncertainties
and a rotation phase should be well described by
a simple $\nu,\,\dot\nu$ spin-down model.
In actuality, most ordinary pulsars show significant deviations
from a simple spin-down model. An analysis of timing noise for
these pulsars has shown that these deviations are due to a random,
continuous wandering of the pulse phase that produces long-term
cubic polynomial components in the timing residuals. These
components lead to the large observed $\ddot\nu$ values and
accordingly to anomalously high values of the braking indices
\citep{cor80,gul82,cor85}.

The properties of timing noise have been studied by
numerous authors over the past 40 yr
\citep{boy72,hel80,cor80,gul82,cor85,arz94,dal95,lyn96,
bay99,hob04,hob10}.
It has been established that timing noise is widespread in 
pulsars. The strength of the timing noise is correlated with
the period derivative and is quantified by a parameter based
on the non-zero second derivative $\ddot\nu$. In fact,
the observed $\ddot\nu$ value is the measure of timing noise
in pulsars \citep{cor85,arz94,hob10}. It is widely believed that
timing noise can be explained by random fluctuations in the pulsar
rotation rate. A description of timing noise is constructed
in the context of physical models that are based on noise-like
processes.

Important results were recently obtained by \citet{hob10} who
presented a detailed study of the timing behavior of 366 pulsars
observed at the Jodrell Bank
Observatory (JBO) over a long time span of 36 yr. It was shown
that a variety of different structures visible in the timing
residuals can be classified into groups. Out of a sample of 366
pulsars, approximately 37\% pulsars have timing residuals
where noise-like variations dominate, about 36\% pulsars have
residuals where cubic structures with both signs dominate, and the
remaining 27\% of pulsars show more complex structures.
\citet{hob10} pay attention to the necessity of using long
data spans to analyze the timing residuals, as otherwise the
structure of the residuals will
depend on the length of the data span analyzed.
\citet{hob10} discussed in detail the problems that are related
to timing noise and the large values of the measured second
derivatives $\ddot\nu$ in pulsars, and paid attention to
the important properties of timing noise. First, there is a large
anti-correlation between the amount of timing noise and
characteristic age. Pulsars with smaller characteristic ages have
more timing noise than older pulsars. Second, there is a
relation between the signature of timing residuals and
characteristic age: the timing residuals observed for the pulsars
with the smaller characteristic ages exhibit mainly cubic
signatures, whereas the timing residuals seen for older pulsars
have quasi-periodic characteristics. The main conclusion of this paper
is that the timing residuals cannot be explained by the models that
are based on noise-like processes because such models
are not consistent with observations over long time scales.

In this paper, we analyze a 33.5 yr period of pulsar timing
observations made at the Pushchino Radio Astronomy Observatory
(PRAO) between 1978 and 2012. The PRAO data sets for 10 pulsars
that are in common with the Jet Propulsion Laboratory (JPL) sample
were combined with the archival JPL data spanning the years
1968--1983 \citep{dow83,dow86}. This procedure extends the
individual data spans for some pulsars up to 43.5 yr.
The common pulsars are B0329+54, B0823+26, B0919+06, B0950+08,
B1133+16, B1642$-$03, B1737+13, B1929+10, B2016+28, and B2217+47.
The combined data set for PSR B1642$-$03 includes also the JBO
data collected over the interval 1981--1999 taken from
an earlier paper \citep{sha01}. The data set for
PSR B1822$-$09 includes the Hartebeesthoek Radio Astronomy
Observatory data collected over the interval 1985--1998
taken from a published paper \citep{sha00}.

\begin{figure}
\epsscale{.80}
\plotone{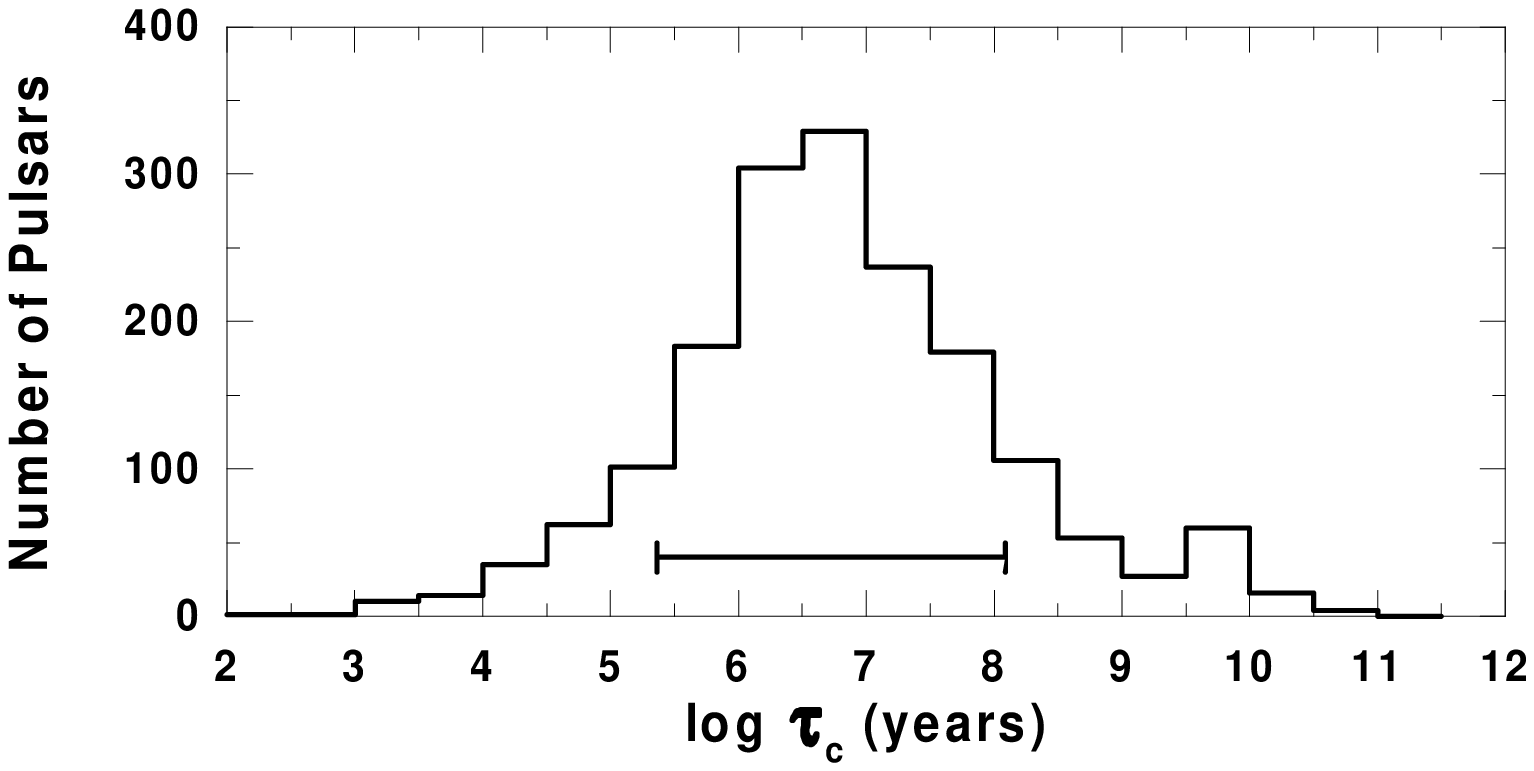}
 \caption{Histogram of characteristic ages of known pulsars
      based on the ATNF Pulsar Catalogue. The age range of the 27
      pulsars from the PRAO sample is indicated by the horizontal
      line. We see that nearly 80\% of the general pulsar population
      has ages within this range.
     \label{gist}}
\end{figure}

While our sample of 27 pulsars is small, these objects
nevertheless reflect the properties of a large number of pulsars.
Figure~\ref{gist} shows a histogram of the characteristic ages
of known pulsars based on the ATNF Pulsar Catalogue \citep{man05}
\footnote {http://www.atnf.csiro.au/research/pulsar/psrcat}.
The 27 studied pulsars have ages ranging from $2.3\times10^{5}$
to $1.2\times10^{8}$ yr; this interval is marked by the horizontal
line. More than 80\% of the global pulsar population has ages
within this range.

On the basis of a detailed analysis of the timing residuals
of these pulsars, we obtained the following results. First,
we detected a new phenomenon in the timing
behavior of two pulsars, B0823+26 and B1929+10, that demonstrates
a rapid change of pulsar rotation parameters in which the sign of
$\ddot\nu$ is reversed. This phenomenon may be interpreted as
a modulation process that changes the sign of $\ddot\nu$.
The second result is that we showed that a long-term structure
of timing noise is produced by four types of rotational
irregularities that have an evolving nature and form three
evolutionary stages in pulsar rotation. According to
the evolutionary scenario, certain types of rotational
irregularities can only occur at certain stages of pulsar
rotation evolution.

\section{Observations and Timing Analysis}

Pulsar timing observations were carried out at the Pushchino
Observatory from 1978 July to 2012 February with a four-year gap
between 1987 and 1990. The observations were made with the BSA
radiotelescope, which operated at frequencies close to 102.7 MHz
until 1998 May and around 111.3 MHz since 1998 November.
A 64-channel radiometer with a channel
bandwidth of 20 kHz was used for the pulsar observations. The data
were sampled at intervals of 2.56 or 1.28 ms. The BSA radiotelescope,
made up of a linearly polarized transit antenna with a beam size of
about (${3.5}/{\cos\delta}$) min, provided the duration of the
observing session ranging from 3 to 11 minutes at different
pulsar declinations $\delta$.
During this time, individual pulses were summed synchronously with
a predicted topocentric pulsar period to form the mean pulse profile
in each 20 kHz channel. After dispersion removal, all the channel
profiles were summed to produce a mean pulse profile for the given
observing session. The topocentric arrival times of the pulses for
each observing session were calculated by cross-correlating the mean
pulse profile with a standard low-noise template. The parameters
of the PRAO timing observations are listed in Table~\ref{inform27}.
In column order, the table lists the pulsar's J2000 and B1950
names, the MJD range of the PRAO observations, the number of pulse
time of arrival (TOA), the mean measurement error for
each observing session, and the dispersion measure (DM) parameter.
The references in the last column mark the DM value that provided
the best adjustment between the two sets of the timing residuals
obtained for each pulsar at the two observing frequencies of
102.7 MHz (until 1998 May) and 111.3 MHz (since 1998 November) in
the PRAO observations.

Timing parameters were determined using the program TEMPO \footnote
{http://www.atnf.csiro.au/research/pulsar/tempo} and the JPL DE200
ephemeris. The topocentric arrival times collected at the PRAO and
the geocentric arrival times obtained from the archival JPL timing
data were all corrected to the barycenter of the solar system
at infinite frequency. The position and the proper motion required
for this correction were taken from \citet{hob04} and were held
constant during the fitting procedure.
The positions of most of the pulsars from our
sample do not require correction since their residuals
(defined as the observed minus predicted arrival times) obtained
after the fitting procedure did not show a sinusoid with a period
of 1 yr within the precision of the measurements.
Position corrections were needed for pulsars B1508+55,
B2020+28, and B2224+65, which are fast moving pulsars,
but a timing solution did not provide improved positions for
these pulsars because of the large timing noise. As the timing
behavior of these pulsars was analyzed over a long interval,
the presence of a weak annual sine wave at the end of this
interval does not affect the large-scale structure of the timing
residuals.

The regular change of pulsar's rotation frequency with time is well
described by a polynomial including several frequency derivatives
\citep{man77}. In accordance with a Taylor expansion, the pulse phase
$\varphi$ at the barycentric arrival time $t$ is expressed as:
\begin{equation}
\varphi(t) = \varphi_{0} + \nu(t-t_{0}) + {\dot\nu}(t-t_{0})^{2}/2 +
{\ddot\nu}(t-t_{0})^{3}/6...,
\end{equation}
where $\varphi_{0}$, $\nu$, $\dot{\nu}$, and $\ddot{\nu}$ are the
pulse phase, rotation frequency, and the first and second frequency
derivative at some reference time $t_{0}$, respectively.
The timing residuals, obtained as the differences between the observed
times and the times predicted by a best-fit model, are used
for the analysis of the rotation behavior of the pulsar.

In order to study the time changes in a pulsar's rotation frequency
in more detail, we show the plots of the timing residuals together
with the plots of the frequency residuals $\Delta\nu$ relative to
the different timing models.
At first, the timing residuals are calculated relative to
a second-order polynomial, including the mean rotation parameters
$\nu$ and $\dot\nu$ defined over the whole data span.
The corresponding frequency residuals $\Delta\nu$ are obtained as
a result of calculating the parameters $\nu$ and $\dot\nu$ from
the local fits, performed to the pulse arrival times over intervals
from 150 to 300 days. The $\Delta\nu$ residuals are given
relative to the mean parameters $\nu$ and $\dot\nu$ and are
referenced to the first point of each interval.
Next, the timing and frequency residuals are calculated relative
to a third-order polynomial for $\nu,\,\dot\nu$, and $\ddot\nu$
using a similar procedure.

The results of the timing analysis for the observed pulsars are given
in Table~\ref{param27}. The columns list, respectively, the pulsar's
B1950 name, the rotation frequency, $\nu$, the first frequency
derivative, $\dot\nu$, and the second frequency derivative, $\ddot\nu$.
The next four columns give the epoch of the $\nu$ measurement,
the MJD range of the observations, the unweighted rms value
remaining after a second-order fit $\sigma_{2}$, and the unweighted
rms value remaining after a third-order fit $\sigma_{3}$.

\section{Results}
\subsection{The Timing Behavior of 27 Pulsars on Long Time Scales}

\citet{lyn10} found that changes in pulse shape parameters
are correlated with changes in the spin-down parameters of some
pulsars. At the PRAO, the pulsar signals are received with
a linearly polarized meridian telescope. For the majority of pulsars
from our sample, the observed shape of the mean pulse profile
varies from day to day because of either the influence of the Faraday
effect or the lack of a sufficient number of individual pulses
for an observing session. Under such conditions, we cannot
determine our pulse shape changes and repeat a similar analysis
of the pulse shape parameters and the spin-down parameters.
Therefore, the pulse shape changes are not discussed in
the later sections.

The timing behavior of the studied pulsars is shown
in Figures~\ref{resid27}(a)--(h).
The residual plots for each pulsar are shown in two panels.
The timing residuals relative to a simple $\nu,\dot\nu$ spin-down
model are shown in the left-hand panels. The corresponding
frequency residuals $\Delta\nu$ are shown in the right-hand panels.
These plots are arranged in order of increasing characteristic age
of the pulsars, $\tau_{c}={P}/{2\dot{P}}$.
The pulsar's B1950 name and the pulse period (seconds)
are indicated in each left-hand panel. The pulsar age in years
is indicated on a logarithmic scale in each right-hand panel.
The periods of the studied pulsars are between 0.1 and 3 s.

It is clearly seen in Figure~\ref{resid27} that the timing behavior
of most pulsars in our sample significantly deviates from a simple
$\nu$, $\dot\nu$ spin-down model. Such deviations, known as timing
noise, may be a result of significant irregularities in the rotation
rate of the pulsars. Table~\ref{param27}
shows that the timing residuals for all the observed pulsars are
characterized by significant second derivatives that are too
large to be attributed to pulsar spin-down. These values of $\ddot\nu$
testify that all our pulsars have a high level of timing noise.

Detailed inspection of the timing residuals plotted in
Figure~\ref{resid27} shows that there is an appreciable correlation
between the amplitude of the residual curve and the characteristic
age of the pulsar---the smaller the amplitude of the residual
curve, the older the pulsar.
As seen in Figure~\ref{resid27}(a), the timing
residuals for younger pulsars have complex structures including
cubic polynomial components with large amplitudes.
As the age of a pulsar increases, we more often observe
residuals that exhibit a quasi-periodic structure with
smaller amplitudes, as is shown in Figure~\ref{resid27}(c)
for PSR B1642$-$03 and B1133+16, for example.
The older pulsars mainly have the structures where
noise-like variations dominate, as in Figure~\ref{resid27}(g),
for example, for the pulsars B0320+39 and B0809+74.

In order to study the effect of timing noise on the $\ddot\nu$
measurements, we analyze the timing
residuals together with the frequency residuals and  try
to determine the causes of the large values of the measured
$\ddot\nu$ for each given pulsar. We distributed all the observed
pulsars into four groups according to the signatures of their
timing residuals. The first group includes eight pulsars,
in which the timing residuals take the form of a cubic
polynomial. Five of these pulsars
have experienced glitches in the past according to the
literature. These pulsars have the largest $\ddot\nu$ values,
as shown in Table~\ref{param27}. The second group includes seven
pulsars that show quasi-periodic variations in their timing residuals.
The third group includes 10 pulsars that exhibit timing
residuals where noise-like variations dominate. The fourth
and final group is unique. This group includes two pulsars,
B0823+26 and B1929+10, whose timing behavior demonstrates
a rare phenomenon---a change in the sign of the second
derivative $\ddot\nu$. Studying this phenomenon will help us
to understand why pulsars with characteristic ages greater
than ${\tau_{c}} > 10^{5}$ ys have both
positive and negative $\ddot\nu$ values.

\subsubsection{The $\ddot\nu$ Values and Cubic Components in
               the Timing Residuals}
The first group includes eight pulsars that show cubic components
in their timing residuals. The amplitude of the timing residuals for
these pulsars varies from one quarter up to more than one pulse
period over spans ranging from 19 to 34 yr. The characteristic ages
of these pulsars range from $2\times 10^{5}$ to $9\times 10^{6}$ yr.

{\bf{(1) B1822$-$09}}.

\begin{figure}
\epsscale{.80}
\plotone{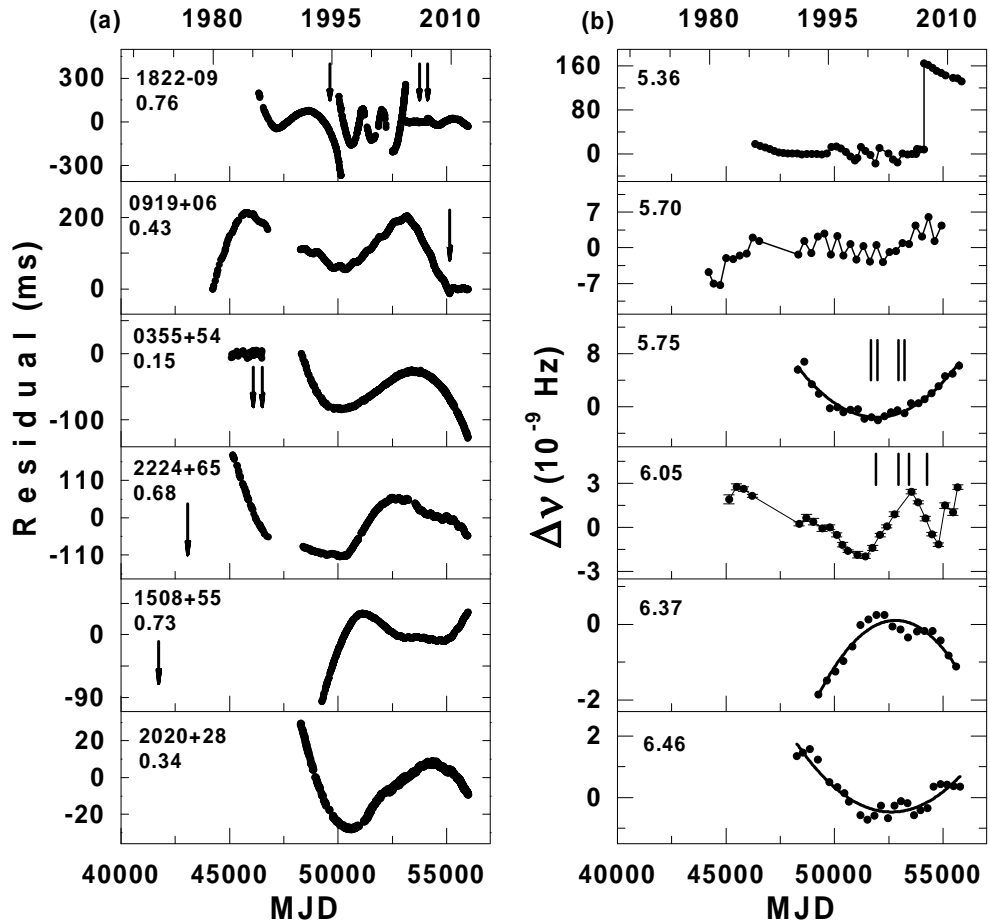}
 \caption{Timing behavior of 27 pulsars between 1968 and 2012.
     (a and b) The left-hand panel shows the timing residuals
     relative to a simple $\nu,\,\dot\nu$ spin-down model. The glitch
     epochs are marked by arrows. The right-hand panel shows the
     corresponding frequency residuals $\Delta\nu$. The epochs of
     micro-glitches with a size of
     ${\Delta\nu}/{\nu}\sim {10^{-10}}$ to $10^{-11}$ are marked
     by vertical lines. A parabolic solid
     curve marks the best fit with a parabolic function. All the
     plots are arranged in order of increasing characteristic age of
     the pulsars. The pulsar B1950 name and the pulse period (seconds)
     are indicated in each left-hand panel. Pulsar age (years) is
     indicated on a logarithmic scale in each right-hand panel.
     Error bars in the $\Delta\nu$ residuals are shown in those panels,
     when they are larger than the plotted points.
     \label{resid27}}
\end{figure}

\citet{sha98} was the first to detect slow glitches in
the rotation frequency of the pulsar B1822$-$09.
Slow glitches represent a unique glitch phenomenon that
results in regular oscillations in the rotation frequency.
A full pattern of the timing and frequency residuals for the
pulsar B1822$-$09 over the 26 yr span between 1985 and
2012 is shown in Figures~\ref{resid27}(a) and ~\ref{resid27}(b),
respectively. These plots show that the pulsar
experienced three discrete glitches (marked
by the three arrows in the left-hand plot) and suffered a sequence
of five slow glitches that occurred over the 1995--2004 interval.
An oscillatory character of the timing and frequency residuals
seen in both the plots over the interval 1995--2004 is due to
the presence of slow glitches. The signature of the large
discrete glitch that occurred in 2007 January \citep{sha09a}
is clearly visible on the right side of the $\Delta\nu$ plot.

The difference between the signature of a slow glitch and that of
a discrete glitch is clearly seen in Figure~\ref{freqnew1822}, where
the time behavior of the frequency derivative $\dot\nu$ and
the $\Delta\nu$ residuals are presented relative to the timing model
1991--1994 throughout the whole time span from 1985 to 2012.
Here, the epochs of the three discrete glitches of 1994, 2006, and
2007 are marked by arrows pointing upward and the epochs at which
the slow glitches occurred are marked by arrows pointing downward.
It is seen in Figure~\ref{freqnew1822}(b) that while the discrete
glitch of 2007 represents a sudden jump in the $\Delta\nu$ residuals,
the slow glitches exhibit gradual exponential increases in
the rotation frequency with a time scale of 200--300 days.
Figure~\ref{freqnew1822}(a) displays the time behavior of $\dot\nu$,
which has some surprising features. First, a gradual
increase in $\Delta\nu$ is accompanied by a rapid
decrease in the magnitude of $\dot\nu$ by $\sim$1\%--2\% of
the initial value and a subsequent exponential increase back
to its initial value. Second, the peaks of $\Delta\dot\nu$,
which characterize the steepness of the front in the residuals
$\Delta\nu$, lie on a parabolic curve that is the envelope
of these peaks. The existence of the envelope indicates
that all of the slow glitches are components of one process
that acted during the 10 yr between 1995 and 2004.
The glitch history of PSR B1822$-$09 was described in detail in
a series of papers \citep{sha98,sha05,sha07,sha09a,sha00}.
The glitch parameters of this pulsar from the PRAO measurements
are given in Table~\ref{glit1822}.

The final part of the oscillatory process over the interval 2000
January--2003 November in PSR B1822$-$09 was also observed by
\citet{zou04}. The authors have independently detected a glitch
event with an unusual signature that occurred over this interval and
have interpreted this glitch in terms of a slow glitch, as
suggested by \citet{sha98}. In a sequence of five slow glitches,
shown in Figure~\ref{freqnew1822}, this glitch is designated as
the third one. The fourth glitch is very small and its signature
is not seen. The glitch events in this pulsar were studied in a
number of other papers \citep{hob10,yua10,esp11,yu13}. Examining
these papers shows that the glitch epochs, indicated in our
Table~\ref{glit1822}, are generally in reasonable agreement with
the measurements of other authors, but the interpretations of the
glitch events are different. According to Table 1 of
\citet{esp11}, PSR B1822$-$09 has experienced six glitches between
MJD 49940 and 54115 and none of these glitches was interpreted
as a slow glitch. We cannot compare a full pattern of the PRAO
glitch residuals, presented in Figure~\ref{freqnew1822}(b), with
that of the JBO glitch residuals because Figures 6 and 7 of \citet{esp11}
show separate residual plots for four different glitches.
The idea of slow glitches has recently received support.
\citet{yu13} suggested that the slow glitch-like events
that result in the quasi-sinusoidal features in the timing
residuals should be labelled as slow glitches.

Figure~\ref{postglt1822} displays the timing residuals for PSR
B1822$-$09 after the glitch of 2007 relative to a simple
spin-down model. The observed cubic term points to a large
positive value $\ddot\nu = 14.4\times 10^{-25}$ s$^{-3}$ that
can be explained by the recovery from the discrete glitch.
We conclude that rotational irregularities, such as slow glitches
and discrete glitches, are the cause of the deviation of
the timing behavior of this pulsar from a simple
$\nu,\,\dot\nu$ spin-down model.

{\bf{(2) B0919+06}}.
As reported by \citet{sha10}, the slow glitch phenomenon was
also detected in the rotation frequency of the pulsar B0919+06.
Figure~\ref{resid27}(a) shows the timing residuals
relative to a timing model of $\nu,\,\dot\nu$, and $\ddot\nu$.
It is seen that the residual curve displays a large quartic term
with the amplitude approximately equal to half the pulsar
period. Rapid oscillations that are imposed on this curve
are interpreted as a sequence of slow glitches.
The arrow in Figure~\ref{resid27}(a) marks the epoch of a large
glitch of magnitude ${{\Delta\nu}/{\nu}}\sim 1.3\times10^{-6}$
that occurred in 2009 November 5 (MJD 55140; \citet{sha10}).
The cubic component of this complex structure corresponds
to a large value $\ddot\nu \sim 2\times 10^{-25}$ s$^{-3}$
which leads to a large braking index, $n \sim 80$.

A sawtooth-like character of the frequency residuals
$\Delta\nu$, shown in Figure~\ref{resid27}(b), implies that
the rotation frequency of this pulsar underwent rapid
oscillations with a time scale of about 600 days during
the whole 30 yr period of observations.
As discussed by \citet{sha10}, the cause of these oscillations
lies in the continuous generation of similar slow glitches.
These glitches are characterized by
small absolute amplitudes of $3.5\times10^{-9}$ Hz, long rise
times of 200 days, and relaxation time intervals of 400 days.
The sequence of the observed slow glitches is well approximated
by a periodic sawtooth-like function with a period of 600 days.
Further observations of this pulsar are needed to study
the relation between the discrete glitch and a process that
continuously generates rapid oscillations identified with slow
glitches. We conclude that rotational irregularities, such as
long-term quasi-periodic oscillations and a sequence of slow glitches
and a large discrete glitch, are responsible for
the large measured $\ddot\nu$ values in this pulsar.

{\bf{(3) B0355+54}}. Figure~\ref{resid27}(a) displays the timing
residuals obtained for the pulsar B0355+54 at the PRAO
during the two observational intervals 1982--1986 and 1991--2012.
This pulsar underwent two glitches in the 1980s.
As reported by \citet{lyn87}, the first glitch occurred in 1985
(MJD 46077) and was quite small with
${{\Delta\nu}/{\nu}}\sim 6\times10^{-9}$. The second occurred in 1986
(MJD 46497) and was a giant glitch
with  ${{\Delta\nu}/{\nu}}\sim 4.4\times10^{-6}$. These glitches
were also registered in the PRAO observations \citep{sha90}.
The glitch epochs are marked by two arrows in Figure~\ref{resid27}(a).

Figure~\ref{resid27}(a) shows that the
timing residuals for PSR B0355+54 over the post-glitch interval
1991--2012 take the form of a cubic polynomial and are
characterized by a significant $\ddot\nu$ value. The
plotted cubic trend presents the residuals calculated relative to
a simple $\nu,\,\dot\nu$ timing model 1991-2010 because
all of the pulse arrival times gathered over the entire 1991--2012
interval cannot be described by a simple spin-down model within
half the pulse period because of the presence of a large $\ddot\nu$.
This plot shows that the amplitude of the cubic structure, equal to
$\sim$ 125 ms, exceeds half the pulse period. The observed cubic
trend corresponds to a positive value
$\ddot\nu\sim 1.7\times 10^{-25}$ s$^{-3}$ that gives a braking
index of $\sim$33.
It is evident that the observed cubic structure in the timing
residuals is caused by the recovery from the 1986 glitch.

The corresponding frequency residuals presented
in Figure~\ref{resid27}(b) also indicate the presence of a large
$\ddot\nu$. From this plot, it is seen that all the $\Delta\nu$
points lie on a parabolic curve that is due to the presence
of a large $\ddot\nu$. \citet{jan06} studied the timing behavior
of this pulsar during $\sim$ 6 yr between 1999 and 2005 and
found four micro-glitches with sizes of ${\Delta\nu}/{\nu}
\sim 0.4, 0.3, 0.4$, and $1.0\times{10^{-10}}$, respectively.
The epochs of these events are marked by vertical lines in
Figure~\ref{resid27}(b). It is seen that these small glitch events
are consistent with the level of fluctuations in the $\Delta\nu$
curve. Therefore, they do not influence large-scale structure in
the frequency residuals and they do not contribute to the measured
value of $\ddot\nu$.
We cannot compare our results with the JBO measurements because
PSR B0355+54 is not included in the JBO sample \citep{hob10}.

The timing and frequency residuals for PSR B0355+54 after a fit
of a third-order polynomial for $\nu,\,\dot\nu$, and $\ddot\nu$
are presented in Figure~\ref{cubic6}. It is seen that the amplitude
of the remaining residuals has considerably decreased.
The $\Delta\nu$ curve is straightened. This plot, as well as
Figure~\ref{resid27}(b), shows that four micro-glitches, whose
epochs are marked by the vertical lines, are local events and
do not influence the measured value of $\ddot\nu$.
In Figure~\ref{cubic6}, it is evident that the second derivative
$\ddot\nu$ dominates the timing and frequency residuals over
the whole span of observations of this pulsar.
Thus, a large cubic trend visible in the timing residuals
of PSR B0355+54 is caused by the recovery from the 1986 glitch.

{\bf{(4) B2224+65}}. The timing residuals of PSR B2224+65 over
the interval 1982--2012 are presented in
Figure~\ref{resid27}(a). This pulsar suffered a large glitch with
a magnitude of ${{\Delta\nu}/{\nu}}\sim 1.7\times10^{-6}$ in 1976
\citep{bac82}. The glitch epoch is marked by an arrow in this plot.
The plot shows that the timing residuals  have a complex
structure with a large amplitude of about 270 ms. The cubic component
of this structure is characterized by a positive $\ddot\nu$,
equal to $\sim 3\times 10^{-26}$ s$^{-3}$, that leads to a braking
index of 106.
The corresponding frequency residuals $\Delta\nu$
are presented in Figure~\ref{resid27}(b). We pay specific attention
to an arc-like extended detail clearly seen in the $\Delta\nu$
curve between 1999 and 2007 ($\sim$ MJD 51200--54100). Four
vertical lines plotted on the $\Delta\nu$ panel mark the epochs
of three micro-glitches with magnitudes of ${{\Delta\nu}/{\nu}}
\sim1.4, 0.8$, and $1.9\times10^{-10}$, respectively, reported
by \citet{jan06}, and the epoch of a fourth glitch with
${\Delta\nu}/{\nu}\sim 3.6\times10^{-10}$, reported by \citet{yua10}.
The plot shows that all of these lines are in the range of
the arc-like extended detail. Most likely, these small glitches
are a consequence of a complex structure of the $\Delta\nu$ curve.

The pulsar B2224+65 is associated with the Guitar Nebula and is
the fastest known pulsar. It has a large transverse velocity of
800--1600 km s$^{-1}$ for a distance of 1--2 kpc. The motion of
PSR B2224+65 through a medium with significant density fluctuations
can cause changes in the DM that can be measured
in timing observations \citep{cha04}. We suppose that the arc-like
extended detail in $\Delta\nu$ can be an important indication of
the existence of DM variations for the pulsar B2224+65.

There are a number of arguments in favor of the DM variations.
First, there is a discrepancy among the DM values measured by
different authors at different epochs. As follows from
Table~\ref{DM}, the difference among the measured DM values
significantly exceeds the estimated measurement uncertainties. At
the PRAO, the DM value was measured using the phase shift between
the two sets of the timing residuals obtained at the two observing
frequencies of 102.746 (until 1998 May) and 111.646 MHz (after
1998 November). The timing residuals for PSR 2224+65, plotted in
Figure~\ref{resid27}(a), were calculated for the value
DM=36.226(5) pc cm$^{-3}$. Secondly, the JBO residuals,
plotted in Figure 3 of \citet{hob10}, show a flatter structure
compared with that of the PRAO residuals, plotted in
Figure~\ref{resid27}(a). The low-frequency timing measurements
are very sensitive to changes in the DM. For example, a continuous
increase in the DM by $\Delta$DM $\sim$ 0.2 pc cm$^{-3}$ during
several years of observations will cause delay in the pulse arrival
times registered at 112 MHz during this interval, where the
delays can be as large as 66 ms. It accordingly follows that
changes of the DM changes with time, if they exist, will more
considerably influence the form of the residuals obtained at 
lower frequencies, for example, 112 MHz. Besides, as is shown
in Figure~\ref{cubic6}, where the timing and frequency residuals
are presented after a fit of a third-order polynomial, a remnant
structure of the arc-like extended detail has the same shape,
as in Figures~\ref{resid27}(a) and~\ref{resid27}(b). This result
means that the $\Delta\nu$ variations over the interval
1999--2007 cannot be attributed to the presence of $\ddot\nu$.
Reliable information about the DM variations can be obtained
from a comparison of the high- and low-frequency timing data
collected over the period 1982--2012 in the quasi-simultaneous
observations of this pulsar at different observatories.

Figure~\ref{cubic6} also shows that the micro-glitches, whose
epochs are marked by the vertical lines in the $\Delta\nu$ plot,
do not influence the large-scale structure of the frequency residuals.
As in the case of PSR B0355+54, small glitch events are local
events and do not contribute to the long-term behavior of $\ddot\nu$.
We conclude that a large positive value of $\ddot\nu$ measured
for the pulsar B2224+65 may be explained by the recovery from a large
glitch in 1976. It is also very likely that the DM variations
additionally contribute to the measured $\ddot\nu$ value.

{\bf{(5) B1508+55}}. The pulsar B1508+55 experienced a small
glitch of ${{\Delta\nu}/{\nu}}\sim 2.2\times10^{-10}$ that
occurred in early 1973 \citep{man74}. Figure~\ref{resid27}(a)
displays the timing residuals for 18 yr of observations after
1993. The epoch of the 1973 glitch is marked by an arrow.
The plotted timing residuals take the form of a cubic polynomial
that is characterized by the negative value
$\ddot\nu =-3.5\times 10^{-26}$ s$^{-3}$; this result leads
to a braking index of -562.
Figure~\ref{resid27}(b) shows the time behavior of the frequency
residuals $\Delta\nu$. From this plot, it is seen that all
the $\Delta\nu$ points are well described by a parabolic curve
despite noticeable quasi-sinusoidal deviations
of these points from the approximating curve. A parabolic pattern
of the $\Delta\nu$ changes means that the second derivative
$\ddot\nu$ dominates the timing residuals during the entire
observational interval.
A quasi-sinusoidal structure in the rotation rate is clearly visible
in Figure~\ref{cubic6}, where the timing and frequency
residuals are shown after the removal of a cubic polynomial.
It is seen that the post-fit timing residuals show two cycles
of a quasi-periodic structure with a variable amplitude of about
30--10 ms. The corresponding $\Delta\nu$ residuals also exhibit
noticeable quasi-periodic variations in the rotation frequency.

Comparing with the results of \citet{hob10} shows that the signature
of the timing residuals, presented in our Figure~\ref{resid27}(a),
is unlike that of the residuals, shown in their Figure 3.
According to Table 1 of \citet{hob10}, this pulsar was observed
at the JBO over the 1982--2006 interval. At the PRAO, B1508+55
was observed at another interval, between 1993 and 2012.
As is shown in Figure 5 of \citet{hob10}, the timing residuals
obtained from different section of the entire data span can have
a different form. It is very likely that we observe this effect
for this pulsar.

Thus, we found that the timing residuals of PSR B1508+55
are characterized by a large negative $\ddot\nu$. Its origin is
uncertain. It is unlikely that the complex structure seen in
the residuals of this pulsar can be a consequence of a small
discrete glitch of 1973.

{\bf{(6) B2020+28}}. The pulsar B2020+28 is not known to have
glitched in the past. Nevertheless, the timing residuals seen in
this pulsar have the form of a cubic polynomial corresponding to a
positive $\ddot\nu$. Figures~\ref{resid27}(a) and~(b)
exhibit the timing and frequency residuals that characterize
the rotation of the pulsar
over the 21 yr span from 1991 to 2012. The observed quasi-cubic
trend in the timing residuals corresponds to a large value
$\ddot\nu\sim3\times 10^{-26}$ s$^{-3}$, which gives a braking
index $n\sim330$. Figure~\ref{resid27}(b) clearly displays that
the changes in the $\Delta\nu$ residuals with time are well
approximated by a parabolic curve that indicates the dominant
character of $\ddot\nu$ in the rotation rate of the pulsar.

The timing and frequency residuals after subtraction of a cubic
polynomial from the pulse arrival times are shown in
Figure~\ref{cubic6}. We see that the remaining timing residuals
have cyclical variations with small amplitude of $\sim$ 8 ms.
The corresponding $\Delta\nu$ residuals, plotted on the right panel,
show noticeable quasi-periodic variations that range
between $\pm0.7\times10^{-9}$ Hz.
We cannot compare our results with the JBO measurements because
PSR B2020+28 is not included in the JBO sample \citep{hob10}.

We hypothesize that the cubic form of the timing residuals
corresponding to a large positive derivative $\ddot\nu$ could be
the consequence of an unobserved glitch event that occurred
in B2020+28 before our observations began in 1991.

{\bf{(7) B0943+10}}. Figure~\ref{resid27}(c) shows the timing
residuals for the pulsar B0943+10 over the 30 yr span from 1982
to 2012. We see that the timing residuals are characterized by
a large cubic trend corresponding to a negative $\ddot\nu$.
The plotted residuals are presented relative to the timing model
1982--2008, as not all the pulse arrival times can be described by
a simple spin-down model within half the pulsar period because
of the presence of a large $\ddot\nu$. This plot shows that
the deviations from the simple timing model are very large,
about $\sim$1500 ms, and that they are more than one pulse period.
The cubic trend in the timing residuals corresponds to a negative
value $\ddot\nu \sim -7\times 10^{-26}$ s$^{-3}$.
This value yields a braking index $n\sim-7924$.
The frequency residuals, presented in Figure~\ref{resid27}(d),
show a parabolic change of the $\Delta\nu$ points with time. This
plot confirms that a large $\ddot\nu$ dominates the timing
residuals of this pulsar over the whole observational interval.

The timing residuals after the removal of a cubic term are shown
in the left panel of Figure~\ref{cubic6}. A large quartic term
visible in the remaining timing residuals points to the presence of
a third significant frequency derivative. The corresponding
frequency residuals $\Delta\nu$ are plotted in the right panel
of Figure~\ref{cubic6}. The noticeable downward slope of
the $\Delta\nu$ curve also indicates the presence of a third
frequency derivative.

Comparing our timing residuals, shown in Figures~\ref{resid27}
and~\ref{cubic6}, and the JBO timing
residuals, shown in Figure 3 and~13 of \citet{hob10}, reveals
a discrepancy between the signatures of these residual
curves. The reason for this discrepancy is related to the presence
of the third derivative in the rotation rate of this pulsar and
the different length of the data sets analyzed. The JBO data set
covers the interval 1983--2006 (Table 1 of \citet{hob10});
the PRAO data set spans more time and covers the interval 1982--2012.
The influence of the third derivative becomes noticeable after
2006, as is seen in Figure~\ref{resid27}(c). If we analyze
the timing data over the same interval 1983--2006, then
the discrepancy between the PRAO and JBO residuals disappears.

Thus, we found that the timing residuals for PSR B0943+10
have the form of a cubic polynomial corresponding to a significant
negative second derivative $\ddot\nu$. Most likely, the pulsar
B0943+10 could have had glitches in its rotation frequency in the past.
Then, a large value of $\ddot\nu$ could be the consequence of these
glitch events that occurred prior to our observations. However,
the reason for the negative sign of $\ddot\nu$ is unclear.

{\bf{(8) B1737+17}}. In the literature, there are no indications
that the pulsar B1737+17 glitched in the past, but
Figure~\ref{resid27}(e) shows a clear cubic trend in the timing
residuals that can be considered the signs of a recovery from
a glitch. The plotted timing residuals are presented over
the interval 1978--2012. The first point from the JPL data set
belongs to 1978 December. It is possible that this pulsar suffered
a glitch before the JPL timing observations started.
The cubic term in the timing residuals corresponds to a positive
value $\ddot\nu \sim 2\times 10^{-26}$ s$^{-3}$.
This value gives a braking index of $\sim$ 5181.
The frequency residuals presented in Figure~\ref{resid27}(f)
show that all the $\Delta\nu$ points lie on a parabolic curve.
This result means that a large $\ddot\nu$ dominates the rotation
behavior of this pulsar over the entire interval of the observations.
The timing and frequency residuals after the removal of a cubic
term are given in Figure~\ref{cubic6}. The remaining residuals
show a noticeable long-term quasi-periodic structure.
Comparison with the results of \citet{hob10} shows that
the timing residuals, presented in our Figures~\ref{resid27}(e) and
~\ref{cubic6}, are in good agreement with the residuals
given in their Figures 3 and 13.

We may suppose that a large positive $\ddot\nu$ measured in
the pulsar B1737+13 could be caused by the recovery from
an unobserved glitch that occurred prior to 1978, before the JPL
observations had begun.

{\bf{Summary}}. We found that all eight pulsars whose
timing residuals take the form of a cubic polynomial have
large second frequency derivatives that range between
$\ddot\nu \sim (0.2--20)\times 10^{-25}$ s$^{-3}$.
The long-term structure of these timing residuals is determined
by such rotational irregularities as discrete glitches,
slow glitches, and quasi-periodic oscillations.

\subsubsection{The $\ddot\nu$ Values and Cyclical Signatures in
               the Timing Residuals}
The second group includes seven pulsars that show cyclical or
quasi-periodic signatures in their timing residuals.
The maximum amplitude of the cyclical changes in the timing
residuals for the different pulsars varies from 4\% to 40$\%$ of
their period over the 30--44 yr spans of the observations.
The characteristic ages of these pulsars range from
$3\times 10^{6}$ to $6\times10^{7}$ yr.

{\bf{(1). B2217+47}}. Figure~\ref{resid27}(c) shows the timing
residuals for PSR B2217+47 over the 42 yr time span between
1970 and 2012. From this plot, we see that the timing
residuals have a quasi-periodic structure over the interval
1970 January--2011 October. This structure is characterized by
maxima spacing between cycles of about 10--14 yr and an amplitude
of about 5--10 ms. We detected that this pulsar experienced
a small glitch that occurred in 2011 October 24 (MJD 55858).
The glitch epoch is marked by an arrow. The $\Delta\nu$ residuals,
plotted in Figure~\ref{resid27}(d), show that the glitch has
a small absolute amplitude of $\Delta\nu\sim 2.3\times10^{-9}$ Hz.
This value corresponds to a fractional increase in the rotation
frequency of ${\Delta\nu}/{\nu}\sim 1.3\times10^{-9}$. The glitch
parameters are given in Table~\ref{glt2217}.
The value of the second derivative measured before the glitch
occurred is $\ddot\nu \sim -7\times 10^{-28}$ s$^{-3}$,
which gives a braking index $n\sim -14$.

Figure~\ref{cyclic7} shows the timing and frequency residuals after
the removal of the cubic trend over the interval before the glitch
occurred. The quasi-periodic structure of the post-fit residuals
is clearly seen in these plots. The $\Delta\nu$ curve changes
within a range of $\Delta\nu$  $\pm 0.8\times10^{-9}$ Hz.
Comparison with the results of \citet{hob10} shows that
our residuals, plotted in Figure~\ref{resid27}(c) before the glitch
occurred, are in good agreement with their timing residuals,
given in their Figure 3.

We conclude that the measured negative value of $\ddot\nu$
is associated with the long-term quasi-periodic variation observed
in the timing residuals for PSR B2217+47 over the interval 1970--2011.

{\bf{(2). B1642$-$03}}. \citet{sha09b} reported the detection
of slow glitches in the rotation rate of the pulsar B1642$-$03
and showed that the observed pattern of cyclical timing
residuals is a result of the continuous generation of slow glitches
whose amplitudes are modulated by some periodic large-scale process.
Slow glitches in this pulsar show striking properties that allow us
to predict the epochs and the amplitudes of new glitches in
the pulsar's rotation rate.
The cyclical character of the timing and frequency residuals for
PSR 1642$-$03 is well seen in Figures~\ref{resid27}(c)
and~\ref{resid27}(d), respectively, which present these residuals
over the 43.5 yr time span from 1968 to 2012.
With respect to the results of \citet{sha09b}, we extend
the observational interval by including three years of new
observations.

In \citet{sha09b}, we showed that the cause of the cyclical
oscillations in the timing residuals of PSR B1642$-$03
is the continuous generation of slow glitches. These glitches are
characterized by long rise times of about 500 days and exponential
relaxation over several years after a glitch.
The observed slow glitches have two surprising properties.
First, the amplitude of these glitches and the time interval
until the following glitch obey a strong linear relation---the
larger the glitch amplitude, the larger the time interval to
the next glitch. This property allows us to predict epochs of new
glitches. Secondly, the amplitude of these glitches is modulated
by a periodic large-scale sawtooth-like function. As a result of
such modulation, the glitch amplitudes change discretely in
a strictly certain sequence assigned by the sawtooth-like function.
This property allows us to predict the amplitudes of new glitches.

In our previous paper \citep{sha09b}, we analyzed the properties
of eight slow glitches. According to the predictions of that paper,
the ninth glitch will occur in $\sim$ 2013.
Figures~\ref{resid27}(c) and~(d) show that this
prediction has come true. The maximum of the ninth glitch is clearly
seen in the timing residuals, plotted in Figure~\ref{resid27}(c).
The maximum of the $\Delta\nu$ residuals, plotted in
Figure~\ref{resid27}(d), occurs around MJD 56100. The observed
epoch is slightly earlier than the predicted epoch of MJD 56300.
Nevertheless, these epochs are in agreement within the time
resolution of $\sim$300 days. Further observations are needed
to determine the amplitude of the ninth glitch. We should measure
the relaxation time interval after the ninth glitch. According to
Table 2 of \citet{sha09b}, for the 43 yr modulation scheme
including eight glitches, this information will be obtained at
the end of 2013 or in the beginning of 2014. In this case,
the observed amplitude of the ninth glitch should be equal to
$\Delta{\nu_{g}} \sim 2.4\times 10^{-9}$ Hz.
In the case of the two other modulation schemes with
modulation periods of 53 yr and 60 yr, the relaxation
time interval will be known in the middle of 2015. Then,
the observed amplitude of the ninth glitch should be equal to
$\Delta{\nu_{g}} \sim 3.9\times 10^{-9}$ Hz.
These two schemes will differ starting with the 10th
glitch that should occur in $\sim$ 2018.

The cyclical structure in this pulsar is characterized by
a positive value $\ddot\nu \sim 1.3\times 10^{-27}$ s$^{-3}$
that yields a braking index $n \sim23$.
We suppose that the large measured value of $\ddot\nu$
in PSR B1642$-$03 is caused by the continuous generation of slow
glitches over the entire period of observations 1968--2012.

Another point of view was presented by \citet{lyn10}, who suggested
that the observed oscillations in PSR B1642$-$03 can be accounted
for by magnetospheric regulated dual-state switching.
In Figure~\ref{resid27}(c), it is seen that the predictability
of the cycles in oscillatory behavior of this pulsar testify that
the interpretation of this process by slow glitch events is
preferable.

{\bf{(3). B1133+16}}. The bottom panel of Figure~\ref{resid27}(c)
presents the timing residuals for the pulsar B1133+16 over
the time span of 43 yr between 1969 and 2012. We see that
the timing residuals have a quasi-periodic character with
an amplitude of $\sim$40 ms and maxima spacing of
$\sim$28 yr. These residuals are characterized by a positive
value $\ddot\nu \sim 1\times 10^{-27}$ s$^{-3}$ that
gives a braking index $n\sim131$.
The corresponding frequency residuals $\Delta\nu$, presented
in Figure~\ref{resid27}(d), have a similar quasi-periodic structure.
The rotation frequency changes within a range $\Delta\nu$ of
$\pm 0.4\times10^{-9}$ Hz. The observed quasi-periodic pattern
of the timing residuals is in good agreement with that of
the timing residuals, given in Figure 3 of \citet{hob10}.

The timing and frequency residuals of PSR B1133+16 after
the subtraction of the polynomial for $\nu,\,\dot\nu$ and
$\ddot\nu$ maintain a quasi-periodic structure,
as is seen in Figure~\ref{cyclic7}.
Thus, an appreciable $\ddot\nu$ value, measured for PSR B1133+16,
is associated with the long-term quasi-periodic variation observed
in the timing residuals of this pulsar.

{\bf{(4) B0329+54}}. The timing residuals of PSR B0329+54 are
shown in Figure~\ref{resid27}(e) over the 43.5 yr time span
from 1968 to 2012. We see that the first half of the timing
residuals exhibit a long-term quasi-periodic variation that
is superimposed on a cubic trend with a rather large amplitude.
The parabolic character of the $\Delta\nu$ residuals, which are
plotted in Figure~\ref{resid27}(f), also testifies that a significant
$\ddot\nu$ dominates in the pulsar rotation rate over the entire
interval of observations, despite a noticeable quasi-sinusoidal
structure visible in the first part of the $\Delta\nu$ curve.
The observed timing residuals are characterized by a positive
value $\ddot\nu \sim 3\times 10^{-27}$ s$^{-3}$
that leads to a braking index of $\sim297$.

\citet{dem79} were the first to argue for the presence of a planet
around the pulsar B0329+54. The authors reported the existence of
a quasi-sinusoidal modulation with a period of 3 yr in pulse
arrival times and supposed that a small planet around the pulsar
could cause such a modulation. The planetary interpretation was
supported by \citet{sha95}, who reported the existence of a
quasi-sinusoidal modulation with a period of 16.9 yr in the
timing residuals and showed that a 3 yr periodicity is clearly
seen in the JPL data after removing the main 16.9 yr
periodicity. Further observations by \citet{kon99} showed that
there is no evidence for the presence of two planet-mass bodies on
the 3 yr and 16.9 yr orbits around this pulsar. The authors
concluded that the observed quasi-periodicity in the timing
residuals of this pulsar may be attributed to the variations in
the rotation frequency or free precession of a neutron star.
According to \citet{hob10}, the JBO data set shows no evidence for
either a 3 yr or a 16.9 yr periodicity and therefore there is
no evidence for a planetary companions to this pulsar.

We agree with this conclusion of \cite{hob10}. In
Figure~\ref{cyclic7}, which shows
the residuals for PSR B0329+54 after the removal of
a cubic term from the arrival times, we see that a 16.9 yr
periodicity existed over the interval 1968--1995 and is not
repeated in further observations after 1995. Note that similar
periodicities can be observed in the rotation frequency of
other pulsars. For example, Figure~\ref{sin1541} shows the timing
residuals for two pulsars, B0823+26 and B1541+09, which contain
two cycles of a quasi-periodic structure. In the case of
B1541+09, the two cycles of the oscillatory residual
curve are well described by a sinusoidal function with a period
of $\sim$7 yr. The observed quasi-periodic structure is not
repeated in further observations. This example illustrates how
a quasi-sinusoidal structure in the timing residuals can confuse
the observer if the observational interval of the given pulsar
is not sufficiently long.

It is very likely that the cubic trend seen in the timing residuals of
Figure~\ref{resid27}(e) and the parabolic character of the
corresponding $\Delta\nu$ residuals, presented in
Figure~\ref{resid27}(f), both are the result of
a large glitch that took place in the distant past. In this case,
the measured $\ddot\nu$ value in the pulsar B0329+54
can be an echo of that previous event. In Figure~\ref{resid27}(f),
we see that the presence of cyclical variation in
the rotation rate influences the pulsar timing behavior only
on rather short intervals. We conclude that the large $\ddot\nu$
value of this pulsar is due to the presence of long-term cyclical
variations and probably the influence of an odd glitch event.

{\bf{(5) B0950+08}}. Figure~\ref{resid27}(e) shows the timing
residuals for the pulsar B0950+08 over the 43.5 yr time span from
1968 to 2012. The plotted timing residuals demonstrate a structure
varying with time. This structure is characterized by a negative
value $\ddot\nu \sim-6\times 10^{-27}$ s$^{-3}$ that
gives a braking index $n\sim-1979$.
The $\Delta\nu$ residuals, plotted in Figure~\ref{resid27}(f),
also indicate some changes in the rotation frequency ranging
from 0 to $-3\times10^{-9}$ Hz. The removal of a cubic term from
the arrival times reduces the amplitude of the timing residuals,
as is shown in Figure~\ref{cyclic7}.
Thus, we suppose that a long-term cyclical structure is the cause
of the significant $\ddot\nu$ measured in the timing residuals
of this pulsar.

{\bf{(6) B1541+09}}. The timing and frequency residuals for
PSR B1541+09 are given in Figures~\ref{resid27}(g) and~(h),
respectively, over the 30 yr interval from 1982--2012.
The timing residual plot shows a cyclical structure that
changes its amplitude and period with time.
A quasi-periodic oscillation seen over the interval 1997--2010
was shown on a large scale before, in Figure~\ref{sin1541}.
This plot shows that the two cycles of the observed
quasi-periodic structure are well
described by a sinusoidal function with a period of $\sim$7
yr and an amplitude of 10 ms. This periodic structure is not
repeated in further observations, as follows from
Figure~\ref{resid27}(g).
The frequency residuals, plotted on the right panel, also
exhibit the presence of cyclical changes in $\Delta\nu$ that
are within the range of $\pm1\times10^{-9}$ Hz.
The observed timing residuals are characterized
by a positive value $\ddot\nu \sim 9\times 10^{-28}$ s$^{-3}$
that gives a braking index $\sim 2012$.

Figure~\ref{cyclic7} shows that the timing and frequency residuals
maintain a cyclical structure after the removal of a cubic term
from the arrival times. Comparison with the results of
\citet{hob10} shows that the signature of the PRAO timing
residuals is generally in reasonable agreement with that of the
JBO residuals, plotted in their Figure 3 up to 2006. One cycle of
the quasi-sinusoidal periodicity examined above is present at
the very end of the JBO data span. It is evident that the presence
of cyclical changes in the rotation frequency is the cause of the
appreciable value of $\ddot\nu$ in PSR B1541+09.  The resemblance
of the timing residuals to the quasi-sinusoidal structure with a
period of $\sim$7 yr is an interesting feature of the residuals.
This result requires further research on the rotation frequency of
this pulsar.

{\bf{(7) B2016+28}}. Figure~\ref{resid27}(g) displays the timing
residuals for PSR B2016+28 over the 43.5 yr time span between 1968
and 2012. In this plot, we see an appreciable cubic trend with
an amplitude of about 30 ms. The parabolic character of
the frequency residuals, presented in Figure~\ref{resid27}(h), also
indicates the dominant influence of a rather significant $\ddot\nu$
over the entire observational interval. The observed timing
residuals are characterized by a positive value
$\ddot\nu \sim 3\times 10^{-27}$ s$^{-3}$ that gives a braking
index $n\sim21186$. In Figure~\ref{cyclic7} that shows
the timing and frequency residuals after the removal of a cubic
term from the arrival times, we see that the remaining residuals
exhibit a long-term quasi-periodic structure.
The signature of the timing residuals, presented in our
Figure~\ref{resid27}(g), is in agreement with that result, as shown
in Figure 3 of \citet{hob10}.

We suppose that the parabolic character of the $\Delta\nu$ curve
for this pulsar, shown in Figure~\ref{resid27}(h), can be a
consequence of the recovery from an old discrete glitch. As in
the case of PSR B0329+54, the deviation of the timing behavior of
PSR B2016+28 from a simple $\nu,\,\dot\nu$ spin-down model can
be explained by the presence of long-term quasi-periodic variations
and probably the influence of a prior glitch event.

{\bf{Summary}}. We found that all seven pulsars whose timing
residuals show a quasi-periodic signature have
measurable $\ddot\nu$ values that range between
$\ddot\nu \sim (0.7--6)\times 10^{-27}$ s$^{-3}$. The structure
seen in the timing residuals of these pulsars is determined by
such rotational irregularities as small discrete glitches, slow
glitches, long-term quasi-periodic processes, and probably
the influence of prior glitch events.

\subsubsection{The $\ddot\nu$ Values and Noise-like Structure
               of the Timing Residuals}
The third group includes 10 pulsars whose timing residuals have
noise-like variations that dominate at a level of less than 1$\%$
of a pulse period over time spans from 21 to 34 yr.
The characteristic ages of these pulsars are more than
$3\times 10^{6}$ yr.

{\bf{(1) B0834+06}}. The top panel of Figure~\ref{resid27}(c)
displays the timing residuals for PSR B0834+06 over the 34 yr
span between 1978 and 2012. This pulsar has a long period of 1.27
s and a sufficiently large period derivative $6.8\times10^{-15}$,
indicating that it is a relatively young pulsar with
$\tau_{c}\sim 3\times 10^{6}$ yr. The study of its rotation
showed that this pulsar has very stable timing behavior in which
rms residuals are $\sim$300 $\mu$s over the entire time span.
Figure~\ref{resid27}(d) shows that weak variations in the
$\Delta\nu$ residuals range between $\pm0.1\times10^{-9}$ Hz. The
timing residuals of this pulsar are characterized by a positive
value $\ddot\nu \sim1\times 10^{-28}$ s$^{-3}$. This value yields a
braking index $n\sim4$ that is very close to the canonical braking
index $n=3$, expected from the simple magnetic dipole braking.
It is very likely that the timing noise in this pulsar
can be accounted for by magnetic dipole braking. In this case,
a measurement of the deterministic $\ddot\nu$ value will be
possible in the near future.

{\bf{(2) B2110+27}}. Figure~\ref{resid27}(e) exhibits the timing
residuals for this pulsar over the 28 yr time span from 1984
to 2012. The observed residuals take the form
of a cubic polynomial with a very small amplitude of $\sim$4 ms.
Figure~\ref{resid27}(f) shows that variations in the $\Delta\nu$
residuals are rather small and range between $\pm0.3\times10^{-9}$ Hz.
The timing residuals are characterized by a positive value
$\ddot\nu \sim4\times 10^{-28}$ s$^{-3}$ that yields a
braking index $n\sim108$. We see that the observed shape of
the timing residuals resembles a damped cubic trend.

{\bf{(3) B2303+30}}. Figure~\ref{resid27}(e) displays the timing
residuals for PSR B2303+30 over the 21 yr span between 1991 and
2012. A weak cubic trend is visible in these residuals.
The observed variations in the $\Delta\nu$ residuals are within
the range of $\pm0.5\times10^{-9}$ Hz, as is shown in
Figure~\ref{resid27}(f). The timing residuals are characterized
by a negative value
$\ddot\nu \sim -9\times 10^{-28}$ s$^{-3}$ that gives a braking
index $n\sim-449$.

{\bf{(4) B1919+21}}. The timing residuals for PSR B1919+21 are
presented in Figure~\ref{resid27}(e) over the 34 yr data span
from 1978 to 2012. The pulsar shows a weak timing noise in which
rms residuals are $\sim$1 ms. The $\Delta\nu$ residuals, plotted
in Figure~\ref{resid27}(f), exhibit random variations that range
between $\pm0.2\times10^{-9}$ Hz.
The observed timing residuals are characterized by a negative
value $\ddot\nu \sim-5\times 10^{-29}$ s$^{-3}$
that gives a braking index $n\sim-65$.

{\bf{(5) B1839+56}}. The bottom panel of Figure~\ref{resid27}(e)
exhibits the timing residuals of PSR B1839+56 over the 29 yr
data span from 1983 to 2012. The timing residuals take the form of
a cubic polynomial with an appreciable amplitude of about 7 ms,
which is still less than 0.5\% of a pulse period.
The corresponding changes in the $\Delta\nu$ residuals, plotted
in Figure~\ref{resid27}(f), are weak and are within a range
$\Delta\nu$ of $\pm0.3\times10^{-9}$ Hz. We may be observing
the damped oscillations of a more long-term cyclical process.
The timing residuals are characterized by a negative
value $\ddot\nu \sim-6\times 10^{-28}$ s$^{-3}$ that
gives a braking index $n\sim-1152$.

{\bf{(6) B2315+21}}. The top panel of Figure~\ref{resid27}(g) shows
the timing residuals for this pulsar over the 21 yr data span from
1991 to 2012. The timing residuals exhibit a noise-like structure.
The corresponding changes in the $\Delta\nu$ residuals, presented
in Figure~\ref{resid27}(h), range between $\pm0.4\times10^{-9}$ Hz.
The timing residuals are characterized by a negative value
$\ddot\nu \sim-2\times 10^{-28}$ s$^{-3}$,
that gives a braking index $n\sim-496$.

{\bf{(7). B0138+59}}. The timing residuals for this pulsar
are shown in Figure~\ref{resid27}(g) over the 29 yr time span
from 1983 to 2012. A noise-like structure is visible in the timing
residuals. The variations of the $\Delta\nu$ residuals, plotted
in Figure~\ref{resid27}(h), are within a small range $\Delta\nu$
of $\pm0.2\times10^{-9}$ Hz. The timing residuals are characterized
by a positive value $\ddot\nu \sim 2\times 10^{-29}$ s$^{-3}$
that gives a braking index $n\sim 215$.

{\bf{(8) B1821+05}}. Figure~\ref{resid27}(g) displays the timing
residuals for PSR B1821+05 over the 21 yr time span between
1991 and 2012. The timing residuals show a noise-like structure.
Figure~\ref{resid27}(h) shows that the variations of
the $\Delta\nu$ residuals are within a range of
$\pm0.4\times10^{-9}$ Hz.
The timing residuals are characterized by a positive
value $\ddot\nu \sim2\times 10^{-28}$ s$^{-3}$
that gives a braking index $n\sim1825$.

{\bf{(9) B0320+39}}. Figure~\ref{resid27}(g) shows the timing
residuals for this pulsar over the 32 yr time span from 1980 to
2012. Noise-like variations are visible in the timing residuals.
The variations in the $\Delta\nu$ residuals range between
$\pm0.2\times10^{-9}$ Hz, as is shown in Figure~\ref{resid27}(h).
The timing residuals are characterized by a positive value
$\ddot\nu \sim 2\times 10^{-29}$ s$^{-3}$ that gives a braking
index $n\sim1455$.

{\bf{(10). B0809+74}}. The bottom panel of Figure~\ref{resid27}(g)
displays the timing residuals for PSR B0809+74 over the 33 yr
data span from 1979 to 2012. The oldest pulsar in our sample
demonstrates timing residuals with a noise-like structure.
Figure~\ref{resid27}(h) shows that variations in
the $\Delta\nu$ residuals range between $\pm0.3\times10^{-9}$ Hz.
The timing residuals are characterized by a positive value
$\ddot\nu \sim 5\times 10^{-29}$ s$^{-3}$.
Thia value yields a braking index of 3717.

Figure~\ref{quiet10} shows the timing residuals for the 10
pulsars after the removal of a cubic term from the pulse arrival
times. We see that most pulsars in this group exhibit random
variations in the remaining timing residuals.

{\bf{Summary}}.  We found that all 10 pulsars whose timing
residuals are dominated by noise-like variations show
$\ddot\nu$ values that range between
$\ddot\nu \sim (2--10)\times 10^{-29}$ s$^{-3}$.
The measured $\ddot\nu$ values are on the boundary of
the precision of our measurements, which is estimated to be
$\sim 1\times10^{-28}$ s$^{-3}$. We suppose that
the observed $\ddot\nu$ values for these pulsars arise
because of the limited precision of our measurements, which
is due to short observing intervals.
As discussed in \citet{liv05}, hundreds of years of observations
are required to measure $\ddot\nu$ due to deterministic spin-down
law in the middle-aged pulsars.

\subsection{Detection of a Change in the Sign of $\ddot\nu$
            for Pulsars B0823+26 and B1929+10}

The fourth group includes two pulsars whose timing residuals
exhibit more complex structure compared with the previous
groups. The main feature of the rotation of these pulsars is
that their rotation parameters $\nu,\,\dot\nu,\,\ddot\nu$ can
undergo a rapid change. This change is accompanied by a change
in the sign of the second frequency derivative $\ddot\nu$.
As a consequence of this phenomenon, the rotation behavior
of these pulsars cannot be described by one
$\nu,\,\dot\nu,\,\ddot\nu$ timing model within one pulse
period over a long time span of tens of years.
Glitch events have nothing to do with this situation.
We show that two different timing models, each of which includes
$\ddot\nu$ of an opposite sign, are required for the description
of the rotation behavior of these pulsars over the 43 yr span
of data analyzed.
The characteristic age of PSR B0823+26 is $\sim 5\times 10^{6}$
yr and that of B1929+10 is $\sim 3\times 10^{6}$ yr.

\subsubsection{Evidence for a Sudden Change in the Sign of
               $\ddot\nu$ for PSR B0823+26}
Figure~\ref{resid27}(c) displays the timing residuals for
PSR B0823+26 over the 43 yr span from 1969 to 2012.
The combined data set includes the JPL data between 1969 and 1983
and the PRAO data between 1981 and 2012. A detailed analysis of
the data showed that all the pulse arrival times gathered over
the 1969--2012 interval cannot be described by a simple
$\nu,\,\dot\nu$ model within a pulse period because of
the presence of a large $\ddot\nu$. These arrival times also
cannot be described within a pulse period by a timing model
including the second derivative $\ddot\nu$.
The cause of this situation is not connected with glitch events.

In order to demonstrate the complex character of the rotation
behavior of this pulsar over the entire observational period
1969--2012, we plotted the timing residuals relative to
a simple spin-down model 1981--1986, as is shown in
Figure~\ref{resid27}(c). From this plot, we see that
the deviation of the residual curve with respect to a zero line
comprises approximately 1700 ms, more than three pulse periods.
In order to keep track of the pulsar rotation over a long
time span, we eliminated the phase jump in the residual curve
by adding one period when the curve deviated by more than one
pulse period.

The cause of such a deviation becomes more clear if we
consider the frequency residuals shown in
Figure~\ref{resid27}(d). Here, the $\Delta\nu$ residuals are
presented relative to the same spin-down model 1981--1986. We see
that the $\Delta\nu$ curve has a distinct break in 1974
November (MJD 42364). The break separates two regions with
different rotation frequency behaviors and is a cross point
of the two curves having convex and concave shapes. It is known
that such curves are characterized by opposite signs of $\ddot\nu$.
This property means that the break in the frequency residuals
is accompanied by a change in the sign of the second frequency
derivative $\ddot\nu$. Further analysis confirms that the sign
of $\ddot\nu$ is reversed at this point. Our purpose is to show
that: (1) the two timing models, including $\ddot\nu$ of the
opposite signs, are required to describe the rotation behavior of
PSR B0823+26 over the entire 1969--2012 interval, (2) the break in
the $\Delta\nu$ curve causes a change in the sign of the second
derivative $\ddot\nu$, and (3) the break in the $\Delta\nu$ curve
causes a distinct jump in the first derivative $\dot\nu$.

Figure~\ref{sign0823} shows a detailed picture of the timing
behavior of the pulsar B0823+26. The top panel demonstrates
how the entire data set for this
pulsar can be described by two independent timing models,
including $\ddot\nu$ with opposite signs. The first interval
{\bf{(I)}} indicates a maximum interspace from the first point,
in which the data set can be described
by one timing model of $\nu$, $\dot\nu$, and $\ddot\nu$ within
half a pulse period. This interval is 1969--1984 (MJD 40264--45873).
The obtained timing residuals are marked by open circles.
An increase of this interspace to the right at one following
point from the data set already causes a phase jump of more
than a pulse period.

The second interval {\bf{(II)}} indicates a maximum interspace
from the last point of the data set (the rightmost
point on the $X$-axis) to the earlier point in the data set
(the leftmost point on the $X$-axis), in which the data set
can be well described by another timing model of $\nu$,
$\dot\nu$, and $\ddot\nu$ within half a pulse period.
The indicated interval is 1971--2012 (MJD 41279--55973).
The obtained timing residuals are marked by solid circles.
An increase of this interspace to the left at one previous
point from the data set already causes a phase jump of more
than a pulse period. The measured rotation parameters for
these timing models include second derivatives $\ddot\nu$
with opposite signs:
$\ddot\nu=-5.51\times10^{-25}$ s$^{-3}$ over the 1969--1984
interval and
$\ddot\nu=0.55\times10^{-25}$ s$^{-3}$ over the 1971--2012
interval. So, we have showed that two timing models
are required to describe the pulsar's rotation behavior
over the 1969--2012 interval and these models include
$\ddot\nu$ of opposite signs.

The middle panel of Figure~\ref{sign0823} shows how a change in
the sign of $\ddot\nu$ has an effect on the rotation frequency of
the pulsar. The vertical line marks the time of the break in the
frequency residuals which, according to Figure~\ref{resid27}(d),
took place in 1974 November (MJD 42364).
The plotted frequency residuals $\Delta\nu$ are presented
relative to the two $\nu,\,\dot\nu,\,\ddot\nu$ timing models:
1969--1974 (before the break) and 1974--2012 (after the break).
The mean rotation parameters for these timing models are given in
Table~\ref{param27}, from which it can be seen that these models
include second derivatives of $\ddot\nu$ with opposite signs.
At first, we present the $\Delta\nu$ residuals relative
to the first timing model 1969--1974, including $\ddot\nu<0$
(open circles). From this plot, we see that the $\Delta\nu$
residuals correspond well with the
zero line in section A (before the break). Beyond section A,
the $\Delta\nu$ curve sharply deviates with respect to the zero
line. This result means that the values of $\nu$ calculated from
the model 1969--1974 do not correspond to the actual rotation
frequencies existing after the break.
A similar picture is observed for the other model 1974--2012,
including $\ddot\nu>0$ (solid circles). In section B
(after the break), the $\Delta\nu$ residuals correspond well with
the zero line. Ahead of section B, the $\Delta\nu$ curve
abruptly declines downwards with respect to the zero line.

Both of these plots, as well as Figure~\ref{resid27}(d), show that
the point of the break is formed as a result of a sudden change in
the slope of the $\Delta\nu$ curve within the framework of the given
model. The sharp turn in the $\Delta\nu$ curve is the cause of a
change in the sign of the second derivative $\ddot\nu$. The break is
a rapid event as it happens within $\sim$300 days, which is the
time resolution. So, we have shown that the break in the
$\Delta\nu$ curve is accompanied by a change in the sign of the
second frequency derivative $\ddot\nu$.

The bottom panel of Figure~\ref{sign0823} shows the first
frequency derivative $\dot\nu$ as a function of time.
The time behavior of $\dot\nu$ is identical for all the timing
models that were mentioned above. The observed changes in
the $\dot\nu$ behavior are the effect of the changes in
the $\Delta\nu$ curve. This statement is confirmed by a visual
comparison of this $\dot\nu$ plot with the $\Delta\nu$ plot,
presented in Figure~\ref{resid27}(d). From the bottom panel
of Figure~\ref{sign0823}, we see that the observed
appearance of the $\dot\nu$ curve testifies to a change
in the sign of $\ddot\nu$. The long-term behavior of the $\dot\nu$
curve can be described as a fast increase of $|\dot\nu|$ before
the break (section A), a distinct jump in the magnitude of
$\dot\nu$ at the point of the break, and the gradual decrease of
$|\dot\nu|$ after the break (section B).
The break in the $\Delta\nu$ curve produces a distinct jump in
the first derivative $\dot\nu$. The magnitude of the jump in
$\dot\nu$ equals
$\Delta\dot\nu \approx 0.12\times10^{-15}$ s$^{-2}$. It makes up
$\sim 2\%$ of the mean value of $\dot\nu\approx
-6.07\times10^{-15}$ s$^{-2}$. So, we have shown that the break in
the $\Delta\nu$ curve causes a jump in the first derivative
$\dot\nu$.

It should be noted that the jump in $\dot\nu$ for the pulsar
B0823+26 was known earlier. \citet{gul82} detected an event
in $\dot\nu$ in early 1975 (JD 2442500). This time agrees well
with our time of the break in the $\Delta\nu$ curve (JD 2442364).
According to their Figure 11, the size of the jump
equals ${\Delta\dot\nu}/{\dot\nu}\sim 2\times10^{-2}$, which
agrees well with our measurement of the jump in $\dot\nu$.
\citet{cor85} also noticed that B0823+26 exhibits
a large discontinuity in both $P$ and $\dot{P}$ near JD 2442600.
These events are clearly visible in their Figure 6.
Later, the timing noise of PSR B0823+26 was analyzed
by \citet{bay99}. The break in the frequency residuals of this
pulsar is clearly seen in their Figure 1. The time of the break
can be estimated from this plot to be $\sim$ JD 2442400.
This time agrees well with our result.

We pay attention to the short-term quasi-periodic variations seen
in the timing behavior of PSR B0823+26 in all three panels of
Figure~\ref{sign0823}. Two cycles of a quasi-periodic structure
of the timing residuals over the interval 1991--2008 were shown
before on a large scale in Figure~\ref{sin1541}. Short-term
quasi-periodic variations in the behavior of $\dot\nu$, seen
clearly in the bottom panel of Figure~\ref{sign0823}, are the cause
of variations in the $\ddot\nu$ values measured during different
time spans. This plot illustrates why the measured values of
$\ddot\nu$ may be different at different parts of the data span
analyzed, as was earlier remarked by \citet{hob04}
(see their Figure 7). On the contrary, the long-term behavior of
the second derivative $\ddot\nu$ is associated with
a linear trend that is observed in the first derivative $\dot\nu$
over the entire 38 yr time span between 1974 and 2012.

\subsubsection{Evidence for a Sudden Change in the Sign
               of $\ddot\nu$ for PSR B1929+10}
The pulsar B1929+10 is the second pulsar
in our research, after PSR B0823+26, for which the phenomenon of
a change in the sign of the second derivative $\ddot\nu$ was
revealed. Figure~\ref{resid27}(c) shows the timing residuals for
B1929+10 over the 43 yr data span from 1969 to 2012. This
plot includes the JPL data spanning between 1969 and 1982 and
the PRAO data spanning between 1999 and 2012.
A 16 yr data gap seen in the residual curve complicates
the analysis of the rotation frequency for this pulsar.
Nevertheless, there are strong arguments for the existence of
a change in the sign of $\ddot\nu$ for this pulsar.

Figure~\ref{resid27}(c) presents the timing residuals relative
to a simple $\nu,\,\dot\nu$ model 1969--1982. We see that
the residuals have different signatures for the two observational
intervals. In the second interval over 1999--2012, the residual
curve suddenly changed its slope with respect to the zero line.
At the end of this interval, the deviation of the timing residuals
exceeded 1000 ms, more than five pulse periods.
Comparing of this plot with the plot of the timing residuals
presented in Figure 3 of \citet{hob10} shows that there is good
agreement between the picture of the PRAO timing residuals and
that of the JBO residuals despite a 16 yr gap in our data
set. In Figure 3 of \citet{hob10}, we see that glitch events
were absent during a 16 yr gap between 1982 and 1999. Therefore,
the observed change of the slope of the residual curve is not
connected with glitches. The variable character of the timing
residuals is probably due to a sudden change in the conditions of
the pulsar rotation. Estimates show that a sudden change in
the pulsar rotation parameters could have taken place in approximately
1997. The picture of the frequency residuals, plotted in
Figure~\ref{resid27}(d), also indicates a variable behavior of
the rotation frequency. We clearly see that the $\Delta\nu$
residuals, presented relative to the timing model 1969--1982,
show different structures for the two observational intervals
1969--1982 and 1999--2012.

In order to reveal the cause of the dissimilar timing behavior
of PSR B1929+10 over the two intervals, we consider
Figure~\ref{sign1929}. Figure~\ref{sign1929}(a) presents
the timing residuals relative to the simple $\nu,\,\dot\nu$
timing model 1969--1982 (on the left side) and relative to
the simple $\nu,\,\dot\nu$ timing model 1999--2012 (on the right
side). We see that the post-fit residuals
exhibit a weak cyclical structure over the first interval
and a large cubic term over the second interval.
The cubic term indicates a large positive value of $\ddot\nu$.
This result means that the change of conditions in the pulsar
rotation have resulted in a large $\ddot\nu$ value.

In the following two plots, we present evidence that
a different signature of the $\Delta\nu$ residuals over the two
observational intervals is due to a change in the value and
the sign of the second frequency derivative $\ddot\nu$.
Figures~\ref{sign1929}(b) and~(c) display the
frequency residuals $\Delta\nu$ relative to the
$\nu,\,\dot\nu,\,\ddot\nu$ timing models 1969--1982 and 1999--2012,
respectively. The mean rotation parameters for these two models
are given in Table~\ref{param27}, which clearly shows that
the models include $\ddot\nu$ of opposite signs.
Figure~\ref{sign1929}(b) shows that the $\Delta\nu$ residuals
relative to the first timing model have a quasi-periodic character
over the interval 1969--1982 and show a significant deviation with
respect to the zero line over the second interval. A parabolic
form of the $\Delta\nu$ residuals over the 1999--2012 interval
indicates the presence of a large $\ddot\nu$, considerably
exceeding $\ddot\nu$ of the first
interval. Still greater discordance between the calculated and
actual rotation frequencies is observed for another timing model,
plotted in Figure~\ref{sign1929}(c).
The frequency residuals $\Delta\nu$ relative to the second
$\nu,\,\dot\nu,\,\ddot\nu$ model have a noise-like character over
the second interval 1999--2012 and show a very large deviation
with respect to the zero line over the first interval. This result means
that the calculated rotation frequencies $\nu$ cannot be compared to
the actual rotation frequencies existing for the first interval.
Both of these plots support a sudden change in the magnitude of
$\ddot\nu$ between the two observational intervals.

The timing behavior of the first derivative $\dot\nu$ is presented
in Figure~\ref{sign1929}(d). We see that the observed signature
of $\dot\nu$ differs between the two intervals. In the second
interval, the magnitude $|\dot\nu|$ decreases with time,
indicating the presence of a large positive second
derivative $\ddot\nu$. In this plot, the solid line indicates
the best fit with a linear function.

We have presented evidence that the pulsar B1929+10 experienced
a significant change in its rotation parameters between
1982 and 1999. A rapid change of the parameters $\nu,\,\dot\nu$,
and $\ddot\nu$ was accompanied by a change in the sign of $\ddot\nu$.
This event could have taken place around 1997 ($\sim$MJD 50400);
this year is marked by a vertical line in Figure~\ref{sign1929}.

\subsubsection{Scheme of the Time Behavior of $\ddot\nu$
               over the Observational Interval 1968--2012}
The scheme of the time behavior of $\ddot\nu$ for two pulsars,
B0823+26 and B1929+10, is plotted in Figure~\ref{modulff}. As was
shown above, the observed change in the value and the sign of
$\ddot\nu$ is due to a rapid change of the pulsar rotation
parameters $\nu,\,\dot\nu$, and $\ddot\nu$. From this plot,
we see that the time behavior of $\ddot\nu$ for PSR B0823+26
is well described by a rectangular function. For the other pulsar,
B1929+10, the behavior of $\ddot\nu$ is shown by the same
function. The plot suggests that we probably observe
a small fragment of some long-term process that can modulate
the second derivative $\ddot\nu$ in pulsars in such a way that
the sign of $\ddot\nu$ is reversed.
If such a modulation of $\ddot\nu$ is a periodical process,
then the time behavior of $\ddot\nu$ could be described by
the periodic rectangular function over long time scales.
In our observations, the phenomenon of a change in the sign of
$\ddot\nu$ is observed for two pulsars out of the 27 studied.
This result means that the period of possible modulation in $\ddot\nu$
insignificantly exceeds the 43 yr time span of our observations
by two to four times and probably reaches 100--200 yr.

So, we have revealed a phenomenon of a rapid change of the pulsar
rotation parameters $\nu$, $\dot\nu$, and $\ddot\nu$. A rapid
change in the rotation frequency $\nu$ is accompanied by
a distinct jump in $\dot\nu$ and a change in the value and
the sign of $\ddot\nu$. The indication of the existence of such
a phenomenon is that the timing behavior of the pulsar cannot
be described by one $\nu,\,\dot\nu,\,\ddot\nu$ timing model
within one pulse period during a long period of observations 
spanning decades. For this case, two models that include
$\ddot\nu$ values of opposite signs are required for the fitting
procedure. Glitches have nothing to do with this situation.

We suppose that in the future this phenomenon---a rapid change of
pulsar rotation parameters accompanied by a change in the sign
of $\ddot\nu$---will be revealed in the rotation frequency of many
other pulsars. Figure 3 of \citet{hob10} presents the timing
residuals for a sample of 366 pulsars. The large cubic trends
seen in the residuals of many pulsars, presented in this plot,
can have different origins.
Some part of these cubic trends may result from glitch events.
The other part of them may serve as indicators that these pulsars
experienced a rapid change in the conditions their rotation.
The cubic trend visible in the timing residuals of PSR B1929+10,
shown in our Figure~\ref{sign1929}(a) over the interval 1999--2012,
indicates this case.

\subsection{The Evolutionary Scenario of the Occurrence of Rotational
            Irregularities}

Figure~\ref{zavis27} shows the relation between the measured value
of the second frequency derivative $\ddot\nu$ and the characteristic
age $\tau_{c}$ for the 27 studied pulsars. These parameters are listed
in Table~\ref{deriv27}. We see that the plotted
points form three sequences according to the signatures of the
timing residuals observed in these pulsars, as was discussed above.
A detailed inspection of Figure~\ref{zavis27} shows that certain
signatures of the timing residuals are correlated with
certain values of the second derivatives $\ddot\nu$
and certain age ranges. This correlation implies that
pulsar timing irregularities have an evolutionary character.
We suggest that the three plotted sequences, corresponding to
different kinds of signatures in the timing residuals, indicate
the three evolutionary stages of pulsar rotation. In other words,
these sequences show that certain signatures in the timing residuals
can occur only at a certain stage of pulsar rotation evolution.
From this fact, it follows that each evolutionary stage is
characterized by certain types of rotational irregularities.

\subsubsection{The Observed Types of Pulsar Rotational Irregularities}

{\bf{Discrete glitches and long-term quasi-periodic oscillations.}}
The extensive JBO timing observations and observations
of other researchers have shown that phenomena such as discrete
glitches and quasi-periodic processes represent
two main types of rotational irregularities in pulsars
\citep{she96,lyn00,wan00,hob04,hob10,yua10,esp11,yu13}.
Glitches occur as sudden, discrete jumps in a pulsar's rotation
frequency, followed by an exponential recovery to the pre-glitch
value. Cubic trends in the timing residuals corresponding to
the positive values of $\ddot\nu$ arise as a result of relaxation
after a discrete glitch \citep{lyn00}. To date, a total of 422
glitches have been revealed in 138 pulsars \citep{esp11,yu13}.
The second type of rotational irregularities occurs in the form of
long-term quasi-periodic oscillations. In Figure 3 of \citet{hob10},
we see that many pulsars show different quasi-periodic
structures in their timing residuals. Both types of rotational
irregularities can occur in the rotation frequency of one pulsar.
According to \citet{hob10}, the timing residuals of many pulsars
represent a combination of a cubic trend and a quasi-periodic
structure. From our data, this effect is clearly visible in
Figure~\ref{cubic6}. The examples of quasi-periodic changes seen
in the timing residuals of the pulsars in our sample
are shown in Figure~\ref{cyclic7}.

{\bf{Slow glitches.}}
Slow glitch phenomena present the third type of rotational
irregularities. The term "slow glitch" was introduced
by \citet{sha98} to explain a glitch of an unusual, gradual
signature that occurred in PSR B1822$-$09 over the interval
1995--1997. According to the literature, the second slow glitch
in this pulsar was observed by \citet{sha00} and the third
glitch was independently observed by \citet{zou04}
and \citet{sha05}. The oscillatory process in PSR B1822$-$09
was finished in 2004. Slow glitches were also investigated
in detail in the rotation frequency of PSR B1642$-$03 and
B0919+06 \citep{sha09b,sha10}. These results showed that
slow glitch events occur in series; they produce regular
oscillations in a pulsar's rotation frequency that looks like
a continuous sequence of glitch-like events. These oscillations
in the rotation frequency give rise to rapid, regular
oscillations in the timing residuals.
As reported by \citet{yua10}, slow glitches have yet to be
identified for two pulsars, J0631+1036 and B1907+10. An extension
of this work would be to study the properties of these events and
to find evidence that they are glitch events, rather than timing
noise. According to \citet{hob10}, slow glitches are not a unique
phenomenon and they can be caused by the same process that causes
timing noise in pulsars. Another point of view
was suggested by \citet{yu13}. These authors proposed that
it is reasonable to label as slow glitches the observed discrete
events that result in quasi-sinusoidal features in the timing
residuals. This fact means that the idea of slow glitches 
has received confirmation from \citet{yu13}.

Apparently, the phenomenon of slow glitches is not widespread
among pulsars. A sample of 10 pulsars whose timing residuals
demonstrate a regular cyclical structure is shown in Figure 1
of \citet{lyn10}. According to a model proposed by
these authors, these timing data are explained by quasi-periodic
switching in the spin-down rate of these pulsars. As discussed
by \citet{jon12}, the harmonic structure seen in the timing
residuals of these 10 pulsars is consistent with the precession
interpretation. We suppose that most pulsars from this sample
are candidates for having slow glitches in their rotation
frequencies.

It should be noted that regular oscillations were observed
earlier in the rotation rate of the Crab pulsar from JBO
observations over the interval 1982--1989. The period of
a regular component of the timing residuals was estimated to be
$\sim$ 20 months \citep{lyn88} or $568\pm10$ days \citep{sco03}.
\citet{lyn88} claimed that the observed cyclical residuals
with this period were consistent with a physically real
quasi-periodic process. In our opinion, the periodic process
with the 568 days period observed in the Crab pulsar resembles
the phenomenon of slow glitches producing a periodic sawtooth-like
modulation with a period of 600 days in the timing residuals of
the pulsar B0919+06 \citep{sha10}.

It is quite possible that the phenomenon of slow glitches
presents an intermediate position between discrete glitches and
quasi-periodic processes. On the one hand, slow glitch events
show properties that should characterize them as discrete glitch
events. For example, a linear relation between
the amplitudes of glitches and the post-glitch intervals
was revealed both for the X-ray pulsar J0537$-$6910 \citep{mid06}
and for the pulsar B1642$-$03 (see Figure 3 of \citet{sha09b}).
On the other hand, a sequence of slow glitches produces regular
cyclical changes in the timing residuals. Regular oscillations
may be a particular case of more general phenomenon related to
long-term quasi-periodic oscillations observed in the rotation
rate of many pulsars. In any case, the phenomenon of slow glitches
is due to real oscillatory processes rather than random
variations in the pulsar rotation rate.

{\bf{A phenomenon of a sudden change in the sign of the second
derivative $\ddot\nu$.}}
In Section 3.2, we reported the detection of a new phenomenon
that is associated with a change of pulsar rotation parameters
$\nu,\,\dot\nu$, and $\ddot\nu$. These parameters change in such
a way that the sign of $\ddot\nu$ is reversed, as is shown
in Figure~\ref{modulff}. This phenomenon was detected in
the timing behavior of two pulsars, B0823+26 and B1929+10.
It is very likely that a similar rapid change of rotation
parameters can occur in many other pulsars. If such a process
cycles in time, then it can be considered a modulation process.
A rough estimate shows that a change of rotation
parameters can occur in the rotation rate of some pulsars about
once every 100 yr. The sign of $\ddot\nu$ will change to the
opposite sign with the same interval.
This process can be a candidate to represent the fourth
type of rotational irregularities.

\subsubsection{Rotational Irregularities in the First Stage of
               Pulsar Rotation Evolution}

In Figure~\ref{zavis27}, we see that the pulsars that show
cubic components in their timing residuals (marked by solid
circles) can be thought of as being in the first stage of pulsar
rotation evolution. These pulsars have large values of $\ddot\nu$
with both signs and are the youngest pulsars in our sample.
The given sequence of
pulsars also includes two pulsars, B0823+26 and B1929+10
(marked by squares), whose timing behavior is determined
by a rapid change of the rotation parameters and a change in
the sign of $\ddot\nu$. In Figure~\ref{cubic6}, we see
that all these pulsars also have a quasi-periodic structure seen
in their timing residuals after the removal of a cubic trend from
the arrival times. At the first stage, there are also slow glitch
events. As was discussed above, two pulsars from the first group,
B1822$-$09 and B0919+06, show a regular cyclical structure
in their timing residuals which are due to slow glitches.

We conclude that rotational irregularities in the first stage of
pulsar rotation evolution are produced by such phenomena as
discrete glitches, slow glitches, quasi-periodic processes,
and processes of a rapid change of pulsar rotation parameters.

\subsubsection{Rotational Irregularities in the Second Stage of
               Pulsar Rotation Evolution}

Figure~\ref{zavis27} shows that the pulsars that stay in
the second stage of pulsar rotation evolution (marked by star
symbols) show timing residuals that are dominated by
quasi-periodic components. These pulsars are older pulsars,
although their age partially overlaps with the ages of
the pulsars remaining in the first stage. These pulsars also
have considerably smaller measured $\ddot\nu$ values.
The timing residuals of these pulsars exhibit a great variety of
quasi-periodic structures, as is seen in Figures~\ref{resid27}
and~\ref{cyclic7}. Besides quasi-periodic processes,
small discrete glitches and slow glitches can also occur in
the rotation frequency of these pulsars. For example,
the cyclical timing residuals of PSR B1642$-$03 are due to
the presence of slow glitches. As discussed above,
the pulsar B2217+47 experienced a small discrete glitch of
magnitude ${\Delta\nu}/{\nu}\sim 1.3\times10^{-9}$ that occurred
in 2011 October. It is very likely that in the second stage
of rotation evolution, events such as large discrete glitches
already will not occur. The process of rapid changes of pulsar
rotation parameters also disappears.

We conclude that rotational irregularities in the second stage of
pulsar rotation evolution are produced by such phenomena as
quasi-periodic processes, small discrete glitches, and slow
glitches. These processes are the cause of measurable values
of $\ddot\nu$ for these pulsars.

\subsubsection{Rotational Irregularities in the Third Stage of
               Pulsar Rotation Evolution}
The third group (marked by triangle symbols) includes pulsars
whose timing residuals exhibit noise-like variations.
In Figure~\ref{zavis27}, we see that the ages of these
pulsars almost completely overlaps with the ages of the pulsars
remaining in the second stage. This result indicates that pulsars
of the same age can remain in different stages of rotation
evolution.

The pulsars that belong to the third stage of pulsar rotation
evolution have many properties in common
(see Figure~\ref{quiet10}).
The periods of these pulsars usually are more than one second.
The timing residuals mainly are due to measurement errors,
although some pulsars of this group can show a weak
remnant structure, as in the case of PSR B1839+56. 
In Figure~\ref{zavis27}, we see that the measured $\ddot\nu$
values for these pulsars are on the boundary of the precision
of our measurements, which is equal to $1\times10^{-28}$ s$^{-3}$.
The limited span of timing data does not allow us to measure
the deterministic $\ddot\nu$ values, which for most pulsars
from this group should be less than $10^{-29}$ s$^{-3}$.

We conclude that rotational irregularities in the third stage
of pulsar rotation evolution should gradually disappear.
At this stage in the timing residuals, noise-like variations
dominate. It follows that in the future these pulsars
will never be observed to glitch.

\subsubsection{The Age Ranges in the Three Stages of
               Pulsar Rotation Evolution}

An analysis of the data, plotted in Figure~\ref{zavis27},
shows that the age boundaries between different evolutional
stages are indistinct and diffusive. This result implies that
the rotation rates of pulsars of the same age can evolve along
different paths.
In Figure~\ref{zavis27} and Table~\ref{deriv27}, we see
that the middle-aged pulsar B0834+06 is approximately the same
age as the pulsars B2217+47 and B1929+10. In contrast to these
pulsars, B0834+06 stays in the third stage of pulsar
rotation evolution. This example illustrates that pulsars of
the same age can have different evolution rotation histories.

From Figure~\ref{zavis27}, we may estimate the age range
for each evolutionary stage. The younger pulsars from our sample
with characteristic ages between  $10^{5}$ and $10^{7}$ yr
belong to the first evolutionary stage. The second
stage includes the middle-aged pulsars with ages greater than
$\sim 10^{6}$ yr. The third evolutionary stage includes both
the middle-aged and old pulsars but the middle-aged pulsars in
the third stage exhibit much less timing noise than pulsars of
the same age in the second evolutionary stage. The uncertainty
of the age boundaries between different evolutionary stages
complicates the correlation between the amount of timing noise
and pulsar age.

A detailed inspection of Figure~\ref{zavis27} shows
that rotation rates of pulsars do not always pass through all
three evolutionary stages. Some pulsars can stay in the first
stage during their whole lives. On the contrary, some
pulsars can escape the first stage of rotation evolution if
they never experience large glitches. Pulsars that have stable
rotation over their whole lives can at once fall into
the third evolutionary stage. The last feature of rotation
evolution implies that, despite the limited accuracy of our
measurements, the deterministic values of $\ddot\nu$ can be
measured for some middle-aged pulsars that stay in the third
evolutionary stage. Among these pulsars, there will always be
pulsars that did not experience any shocks in their life and
consequently have stable rotation and a low level of timing noise.

For example, we examine the pulsar B0834+06. This middle-aged
pulsar has a relatively low level of timing noise, as discussed
above (see Figure~\ref{resid27}(c)). We may suppose that
neither discrete glitches nor quasi-periodic oscillations
disturbed the rotation of this pulsar during its life.
Figure~\ref{zavis27} shows that
the measured $\ddot\nu$ value for this pulsar falls very
close to the sloping line of the deterministic values
of $\ddot\nu$, indicating that slowdown of the rotation of this
pulsar is closely related to a simple $\nu,\,\dot\nu$ spin-down
model. We conclude that a measurement of the deterministic
value of $\ddot\nu$ for this pulsar is possible in the near
future rather than in hundreds of years as for the other
older pulsars.

\subsubsection{The Magnetic Fields in the Three Stages of
               Pulsar Rotation Evolution}

Evolution of the magnetic field of a neutron star was discussed
in detail in a series of papers
\citep{rud91a,rud91b,rud04,rud06,rud98}.
We consider the evolutionary picture of the magnetic field for
the middle-aged and old pulsars taking into account the signatures
of the timing residuals observed in these pulsars.
Figure~\ref{zavb0}(a) shows the dependence of the surface magnetic
field strength $B_{s}$ on the characteristic age $\tau_{c}$ for
the 27 studied pulsars. The parameters $B_{s}$ and $\tau_{c}$
are listed in Table~\ref{deriv27}.
The three plotted sequences indicate
that the magnetic field values $B_{s}$ considerably differ for
the three groups of pulsars and that these values are furthemore
correlated with certain signatures of the timing residuals.
Here, a sequence of the $B_{s}$ values for the pulsars with cubic
signatures of the timing residuals is marked by solid circles,
pulsars with quasi-periodic signatures in their residuals
are marked by star symbols, and pulsars with noise-like
signatures in their residuals are marked by triangle symbols.

We see two surprising features in the observed correlation picture.
First, the $B_{s}$ values of the pulsars showing noise-like
components in their timing residuals (the $B_{nois}$ values),
is are greater than the $B_{s}$ values of the pulsars of the same
age that show timing residuals dominated by cubic
components (the $B_{(cub)}$ values) or quasi-periodic components
(the $B_{(cyc)}$ values). This result may mean that magnetic
fields are stronger for pulsars whose rotation is more stable.

Second, a sequence of $B_{(cyc)}$ values continues well into
a sequence of the $B_{(cub)}$ values. This result indicates
a common evolutionary path for these two groups of pulsars.
For analysis, these two sequences are combined into one sequence
$B_{(cub+cyc)}$, as is shown in Figure~\ref{zavb0}(b). We see
that the two straight lines fit to the sequences of
$B_{(nois)}$ points and $B_{(cub+cyc)}$ points have identical
slopes. For pulsars of the same age, the $B_{(nois)}$ values
nearly four times greater than the $B_{(cub+cyc)}$ values.
This result suggests that the magnetic fields are approximately
four times stronger for the pulsars that have stable
rotation. The dependence of the $B_{s}$ values on different
kinds of signatures of the timing residuals is a good
indication of the validity of the supposition that the rotation
rates of pulsars pass through different evolutionary stages
and that some pulsars can have a special path of evolution
that is characterized by a very stable pulsar
rotational behavior during the whole pulsar's life. We conclude
that the observed correlation picture of the surface magnetic
field strength $B_{s}$ with certain signatures of the timing
residuals, presented in Figure~\ref{zavb0}, provides convincing
evidence for the existence of the evolutionary scenario of
the origin of rotational irregularities.

In order to check the results obtained above for a large number of
pulsars, we analyzed the JBO timing residuals, presented in Figure
3 of \citet{hob10}. We tried to classify ordinary pulsars of
middle and older age according to the signatures of their
timing residuals. The obtained results are shown in
Figure~\ref{zavjbo}. Figure~\ref{zavjbo}(a) displays a relation
between the measured values of $\ddot\nu$ and the characteristic
age $\tau_{c}$. Here, 73 selected pulsars showing timing
residuals dominated by cubic and quasi-periodic components
are marked by squares and 64 pulsars showing noise-like
components in their timing residuals are marked by open circles.
The plotted three sequences present
the PRAO data, as in Figure~\ref{zavis27}. We see that the
distribution of the plotted points does not contradict the
supposition about the existence of three stages of pulsar
rotation evolution.

Figure~\ref{zavjbo}(b) displays the relation between the surface
magnetic field strength $B_{s}$ and $\tau_{c}$ for the selected
pulsars from the JBO sample. The plotted two sequences present
the PRAO data, as in Figure~\ref{zavb0}(b). In this plot, the
two straight lines fit to two groups of JBO points show
that magnetic fields are stronger for pulsars
exhibiting noise-like signatures in their timing residuals.
This result confirms that pulsars that stay in the third
evolutionary stage have a tendency to show stronger magnetic fields
than pulsars in other stages. We conclude that the evolutionary
scenario of the occurrence of rotational irregularities is
confirmed by the dependencies that are presented in
Figure~\ref{zavjbo} for a large number of pulsars.

\section{Discussion}

\subsection{Explanation of the Large Values of $\ddot\nu$
            with Both Signs}

We suppose that the large values of $\ddot\nu$ with both signs
are the result of the existence of four types of rotational
irregularities that have evolutionary characteristics and form
three evolutionary stages in pulsar rotation rate. Rotational
irregularities alone are responsible for the large values of
the measured second derivative $\ddot\nu$ and the variable
structure of the timing noise over a long time scale.

As discussed above, rotational irregularities occur in
the form of discrete glitches, quasi-periodic oscillatory
processes, slow glitches, and processes that change the pulsar
rotation parameters $\nu,\,\dot\nu$, and $\ddot\nu$ in such
a way that the sign of $\ddot\nu$ is reversed.
An analysis of timing noise
showed that several types of rotational irregularities can
simultaneously occur in the rotation rate of one pulsar.
For example, the rotation frequency of PSR B0919+06 over
the interval 1979--2009 underwent large changes that were due to
three types of rotational irregularities including a series of
19 slow glitches, a long-term quasi-periodic process, and a large
discrete glitch \citep{sha10}. We conclude that rotational
irregularities are the cause of the deviation of the timing
behavior of most pulsars from a simple $\nu,\,\dot\nu$ model.
This conclusion is in accordance with the statement by
\citet{hob10} that the timing noise cannot be explained by
models that are based on random walks in the pulse phase,
frequency, or spin-down rate. Such models are not consistent
with the observations.

It is known that the $\ddot\nu$ values for the simple model
of magnetic dipole braking should be positive. However,
timing observations show that pulsars with characteristic
ages more than $\tau_{c}>10^{5}$ yr have both the positive and
negative $\ddot\nu$ values. As pointed out by \citet{hob10},
approximately 52\% of these pulsars have a positive $\ddot\nu$
value and 48\% have negative $\ddot\nu$. These authors noted that
this situation is not related to glitch recovery nor magnetic
dipole radiation, but is a result of some other process.

We propose that we have found a process that can result in
a change in the sign of $\ddot\nu$ in pulsars.
As was reported in Section 3.2, this process changes the pulsar
rotation parameters together with a change in the sign of $\ddot\nu$,
as shown in Figure~\ref{modulff}. This process was detected in
the timing behavior of two pulsars, B0823+26 and B1929+10.
It is very likely that a similar rapid change of rotation
parameters can occur in many other pulsars. If so, then this
process could explain the problem that approximately
equal numbers of pulsars have timing residuals in the form
of a cubic polynomial with positive (20\%) and negative (16\%)
signs of $\ddot\nu$ \citep{hob10}.
While glitch recovery produces cubic trends corresponding to
the positive values of $\ddot\nu$, this process could cause
large values of $\ddot\nu$ with both signs. It should be noted
that such a process apparently can operate only
in pulsars that stay in the first stage of pulsar rotation
evolution. In the final stages of rotation evolution, negative
signs of $\ddot\nu$ can result from the limited temporal baseline
of data of a more long-term oscillatory process.

In the literature, there are different interpretations of
the origin of large $\ddot\nu$ values. In the paper
by \citet{ura06}, the large  values of $\ddot\nu$ with both signs
are attributed to small variations in the spin-down torque.
\citet{bir12} discussed this problem in the context of the
pulsar spin-down model that involves the existence of a long-term
cyclical process changing the observed rotation parameters. This
cyclical process is superimposed on the secular spin-down of
pulsars and produces cyclical variations in $\ddot\nu$ with
time scales of a few thousand years. The second
interpretation is in accordance  with our experimental
result describing the fourth type of rotational irregularities.
This type of irregularity produces a change in the sign of
$\ddot\nu$ owing to a rapid change of pulsar rotation parameters.
However, we explain the origin of the large $\ddot\nu$ values
and the variable structure of the timing noise by the existence
of four types of rotational irregularities that have
evolutionary characteristics.

\subsection{Explanations for Other Observed Properties
            of Pulsar Rotation}

We propose an evolutionary scenario for the occurrence of rotational
irregularities to explain the observed properties of timing noise
in pulsars. Timing noise is common among pulsars because rotational
irregularities occur in pulsars of all ages, but the strength
of timing noise depends on the evolutionary stage of pulsar rotation,
as is shown in Figure~\ref{zavis27}. Younger pulsars
stay in the first evolutionary stage of pulsar rotation where
the pulsar rotation frequency is subject to all four types of
rotational irregularities---discrete glitches, quasi-periodic
oscillations, slow glitches, and processes that change pulsar rotation
parameters, and, as a consequence, change the sign of $\ddot\nu$.
These pulsars have the largest $\ddot\nu$ values and show strong
timing noise. In their timing residuals a large cubic trend
dominates that is the result of relaxation after a discrete glitch
(for $\ddot\nu>0$) or the result of a process of a change in
the sign of $\ddot\nu$ (for $\ddot\nu$ with both signs).
The middle-aged pulsars stay in the second evolutionary stage
of pulsar rotation and their timing activity is related to three
types of rotational irregularities---quasi-periodic oscillations,
slow glitches, and small discrete glitches. The timing residuals
of such pulsars exhibit mainly a quasi-periodic character.
The strength of the timing noise of old pulsars that stay in
the third evolutionary stage weakens because rotational
irregularities at this stage gradually disappear.
The timing residuals of old pulsars are mainly dominated by
noise-like variations. This presented evolutionary picture
is in accordance with the results of \citet{hob10}, who
discussed the features of timing residuals depending on
the characteristic age of pulsars.

The age boundaries between different evolutionary stages are
diffusive. This feature of pulsar rotation evolution explains
why two pulsars with similar rotation parameters
exhibit different signatures in their timing residuals.
For example, two pulsars, B0943+10 and B1133+16, have almost
identical rotation parameters $\nu$ and $\dot\nu$, as follows
from Table~\ref{param27}, but exhibit different structures in
their timing and frequency residuals, as seen
in Figures~\ref{resid27}(c) and ~\ref{resid27}(d).
This result occurs because these pulsars have different evolutionary
histories; they evolve along different paths and therefore belong
to different evolutionary stages of pulsar rotation.
In Figure~\ref{zavis27}, we see that the pulsar B0943+10
stays in the first evolutional stage, while the pulsar B1133+16
stays in the second stage of pulsar rotation evolution.

It was believed that all pulsars can experience
glitches \citep{alp94,alp06,mel08}. The evolutionary scenario
for the origin of pulsar rotational irregularities is inconsistent
with this statement. Pulsar glitches are not common phenomena
experienced by pulsars of all ages. Figure~\ref{zavis27} shows
that pulsar glitches occur more frequently in younger pulsars as
these pulsars belong to the first evolutionary stage of pulsar
rotation. Pulsar glitches will never occur in old pulsars that
belong to the third evolutionary stage of pulsar rotation. Pulsars
in the final evolutional stage either already have experienced the
stage of glitches in their rotation rate or have never experienced
any shocks in their life and so immediately entered the final
stage of pulsar rotation evolution. As a consequence, one pulsar may
regularly experience glitch events while another pulsar with similar
properties has never been observed to glitch.

\section{Summary}

On the basis of a detailed analysis of the timing behavior
of 27 pulsars over long data baselines spanning up to 43.5 yr,
we have revealed the following important features of pulsar
rotation.

  1. We detected a new type of rotational irregularity that
is the result of a rapid change of pulsar rotation parameters
in such a way that the sign of $\ddot\nu$ is reversed. This process
was revealed in the timing behavior of two pulsars, B0823+26 and
B1929+10. The detected phenomenon of changing the sign of
$\ddot\nu$ can help to explain the problem why pulsars with
characteristic ages older than $\tau_{c}>10^{5}$ yr have
both positive and negative $\ddot\nu$ values.

  2. We showed that a variable structure of timing noise
over long time scales is produced by four types of
rotational irregularities that occur in the form of discrete
glitches, quasi-periodic oscillations, slow glitches, and
a process changing pulsar rotation parameters together
with a change in the sign of $\ddot\nu$.

  3. We found that all four types of observed rotational
irregularities have evolutionary nature and form three
evolutionary stages in pulsar rotation rate. This result means
that a certain type of rotational irregularity can occur only
at a certain stage of pulsar rotation evolution.
In the first stage of rotation evolution,
including the younger pulsars with $\tau_{c}<10^{7}$
yr, all four types of rotational irregularities
occur---discrete glitches, slow glitches, quasi-periodical
oscillations, and a process of changing pulsar rotation parameters
and the sign of $\ddot\nu$. In the second stage of rotation
evolution, including the middle-aged pulsars with
$\tau_{c}>10^{6}$ yr, large glitches and the process that changes
the sign of $\ddot\nu$ disappear. Rotational irregularities include
only small discrete glitches, slow glitches, and quasi-periodical
oscillations. In the third stage of rotation evolution, including
both the middle-aged and old pulsars, rotational irregularities
should gradually weaken and disappear entirely. In this stage,
the timing residuals are mainly dominated by noise-like variations.
The evolutionary nature of rotation irregularities indicates that
the deterministic values of $\ddot\nu$ and accordingly to the true
values of the braking indices $n$ can be never measured for ordinary
pulsars that stay in the first or second stages of pulsar rotation
evolution.

  4. We found that the surface magnetic field strength $B_{s}$ for
pulsars is correlated with certain signatures of the timing
residuals, as shown in Figure~\ref{zavb0}. We revealed two
surprising features in this correlation picture. The first feature
is that the magnetic fields are stronger for pulsars that show
noise-like signatures in their timing residuals, that is, for the
pulsars whose rotation has a more stable character. The second
feature is that the magnetic fields for two groups of the pulsars,
those exhibiting timing residuals with cubic signatures and
those exibiting timing residuals with quasi-periodic signatures,
smoothly vary from one group to another. This result indicates
that these two groups of pulsars have a common evolutionary path.
The observed correlation of the magnetic
field with certain signatures of the timing residuals confirms
the evolutionary nature of rotational irregularities.

  5. We found that the age boundaries between different
evolutionary stages are indistinct and diffusive. This result
is because different pulsars with similar properties evolve
along different paths and have different evolutionary histories.
The first corollary of this feature is that
pulsar glitches are not a common event experienced by pulsars
of all ages. Pulsar glitches will never occur in old pulsars
that belong to the third evolutionary stage of pulsar rotation.
The second corollary of this feature is that the measurements
of the deterministic values of $\ddot\nu$ are possible for
some middle-aged pulsars that stay in the third evolutionary
stage and show stable rotation and a low level of timing noise.
It is very likely that these pulsars did not experience any
shocks over their lifetimes. Then, the measurements of
the deterministic $\ddot\nu$ values for such pulsars will not
require very long, in hundreds of years, data baselines.

  6. We detected a small glitch in the pulsar B2217+47.
The glitch occurred in 2011 October 24 (MJD 55858) and was
characterized by a fractional increase in the rotation
frequency of ${\Delta\nu}/{\nu}\sim 1.3\times10^{-9}$.

  7. We showed that there are arguments in favor of the
existence of continuous changes in the DM
over the interval 1999--2007 for the pulsar B2224+65.

In summary, we have studied the properties of the timing
behavior of 27 pulsars whose characteristic ages range from
$10^{5}$ to $10^{8}$ yr. We propose that the evolutionary
scenario of the occurrence of rotational irregularities can be
generalized to all ordinary pulsars in the given age range.
Very likely, the timing residuals of younger pulsars can also
be well explained in the context of the evolutionary scenario
of the occurrence of rotational irregularities.

\acknowledgments
We are grateful to R. D. Dagkesamansky for discussions and
valuable comments and the staff of the PRAO for their aid in
carrying out the many-year pulsar observations on the BSA
radiotelescope. This work was supported by the Russian
Foundation for Basic Research (grant No. 09-02-00473) and
the Program of the Russian Academy
of Sciences Physical Division "Active Processes in Galactic
and Extragalactic Objects". The authors are grateful to the
referee for helpful comments and suggestions.

\newpage
\clearpage

\newpage
\clearpage
\newpage
\clearpage
\begin{figure}
\epsscale{.80}
\plotone{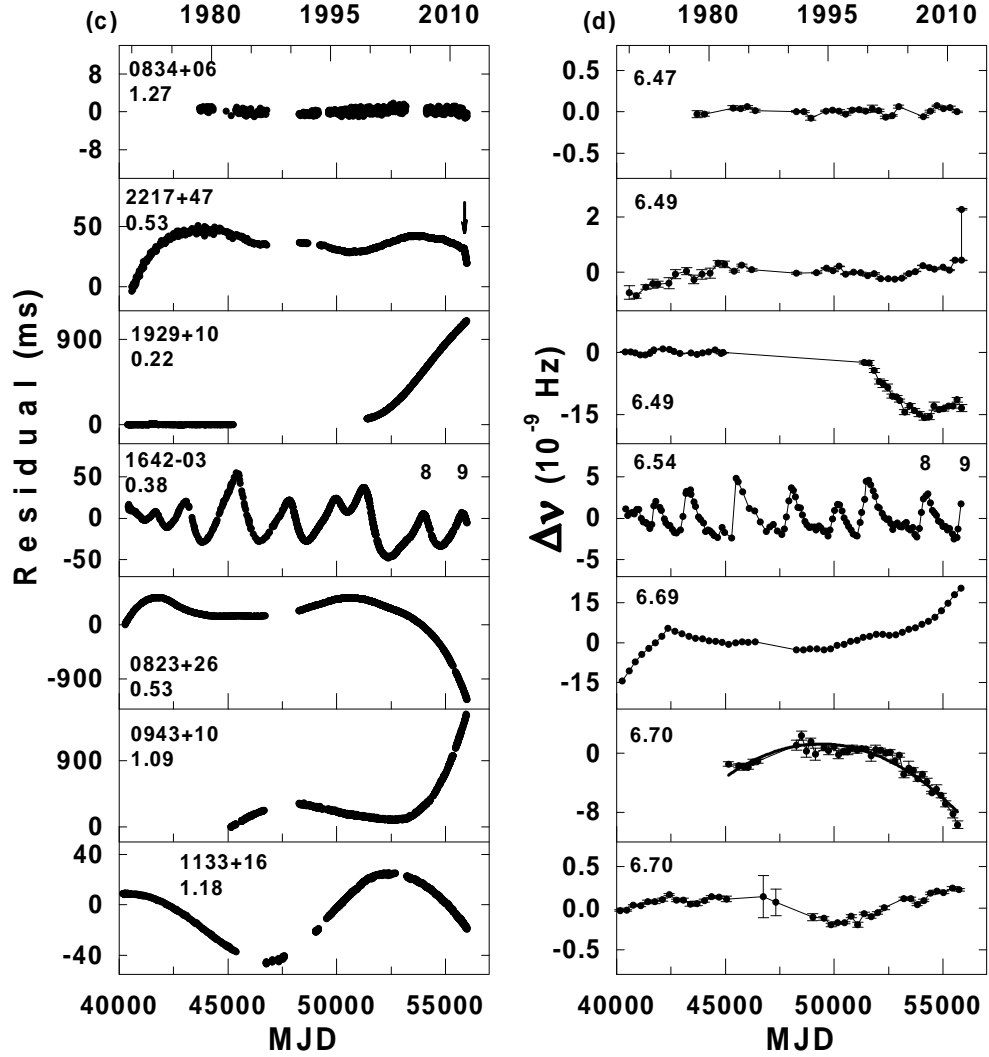}
 \caption{Figure 2-continued. (c and d) In the panel of PSR B2217+47,
    an arrow marks the epoch of the glitch that occurred in 2011 October 24
    (MJD 55858). The corresponding $\Delta\nu$ plot shows that this
    glitch has a small absolute amplitude of
    $\Delta\nu \sim 2.3\times10^{-9}$ Hz.
    In the panel of PSR B1642$-$03, a sequence of slow glitches
    includes eight slow glitches and shows a new, ninth glitch
    seen at the very end of the data set.
     \label{resid27cd}}
\end{figure}

\newpage
\clearpage
\begin{figure}
\epsscale{.80}
\plotone{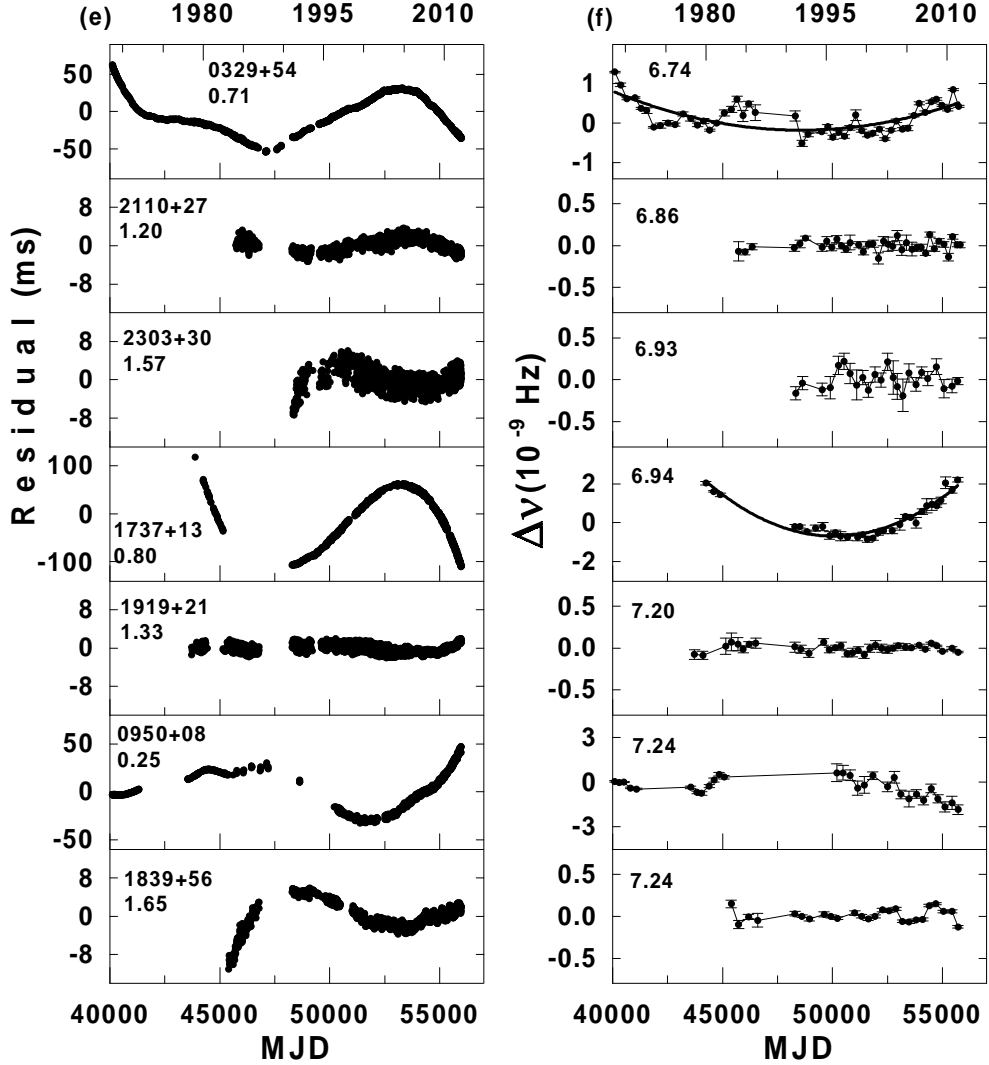}
 \caption{Figure 2-continued. (e and f) In the panel of PSR
 B0329+54, it is seen that the first half of the timing residuals
 contains the long-term cyclical variation that is superimposed
 on an appreciable cubic trend. The parabolic character of the
 corresponding $\Delta\nu$ residuals indicates that
 a significant $\ddot\nu$ dominates in the residuals during
 the whole interval of observations despite a noticeable
 quasi-sinusoidal structure visible in the first
 half of the $\Delta\nu$ curve.
     \label{resid27ef}}
\end{figure}

\newpage
\clearpage
\begin{figure}
\epsscale{.80}
\plotone{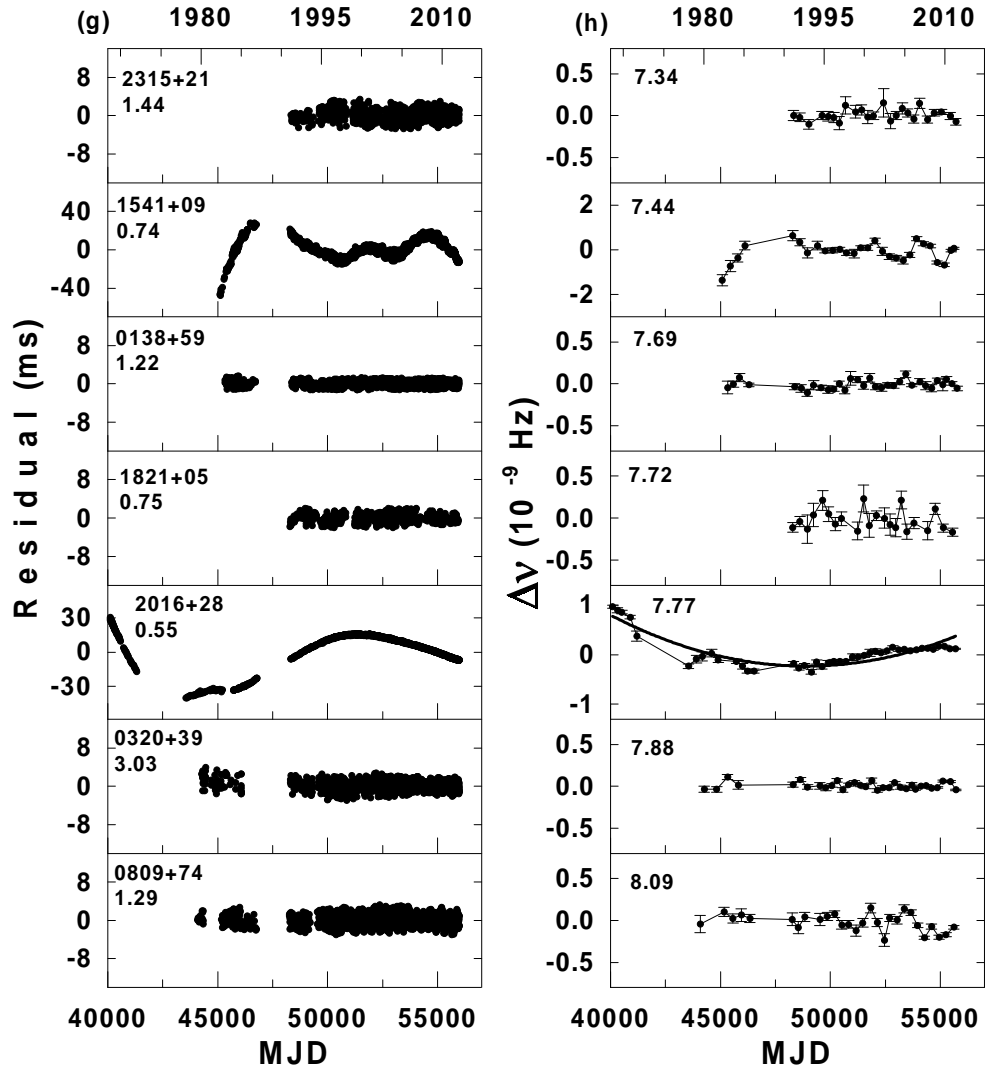}
 \caption{Figure 2-continued. (g and h) In the panel of
      PSR B1541+09, the timing residuals show two cycles of
      a clear quasi-sinusoidal structure visible over the interval
      1995--2011.
     \label{resid27gh}}
\end{figure}

\newpage
\clearpage
\begin{figure}
\epsscale{.80}
\plotone{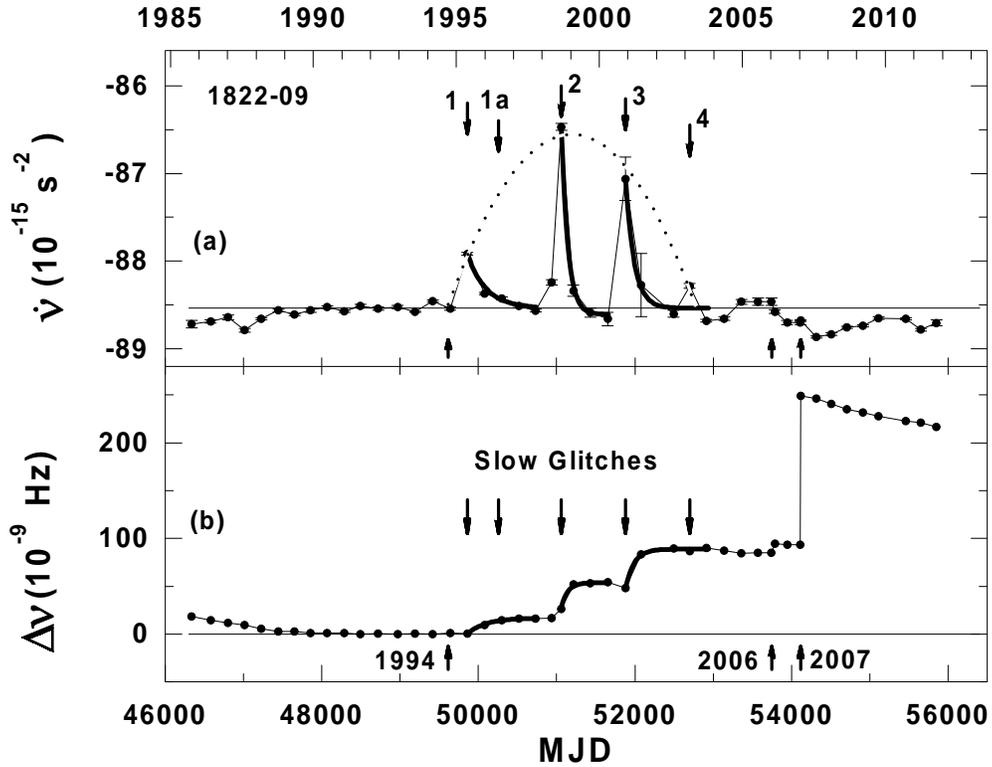}
 \caption{Five slow glitches and three discrete glitches in PSR B1822$-$09.
    Arrows pointing downward indicate the epochs at which the
    slow glitches occurred while arrows pointing upward indicate
    the discrete glitches. (a) The changes of $\dot\nu$ over
    the 1995--2004 interval are due to the slow glitches. We see
    that the $\Delta\dot\nu$ peaks lie on a parabolic curve that is
    the envelope of these peaks. (b) The $\Delta\nu$ residuals relative
    to a simple $\nu,\,\dot\nu$ model 1991--1994. The gradual increase
    in $\Delta\nu$ over the 1995--2004 interval is due to the slow
    glitches. The exponential fits to the $\dot\nu$ curve and
    the $\Delta\nu$ curve are drawn with bold lines in panels (a) and (b).
    The signature of the large glitch of 2007 is clearly seen on the
    right side of the $\Delta\nu$ plot.
     \label{freqnew1822}}
\end{figure}

\newpage
\clearpage
\begin{figure}
\epsscale{.80}
\plotone{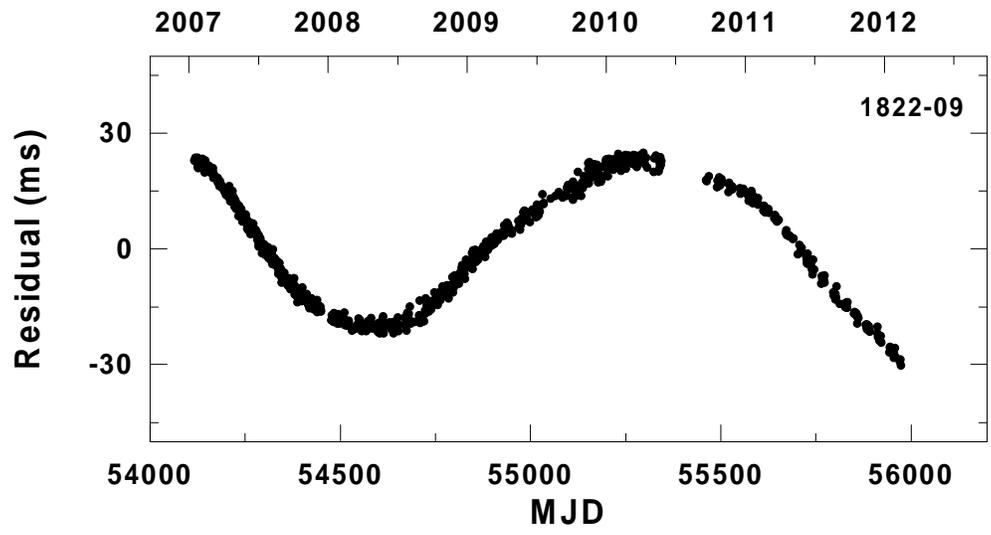}
 \caption{Timing residuals for PSR B1822$-$09 after the glitch
      of 2007 relative to a simple spin-down model. The observed
      cubic term points to a large positive second derivative
      that is explained by the recovery from this glitch.
     \label{postglt1822}}
\end{figure}

\newpage
\clearpage
\begin{figure}
\epsscale{.80}
\plotone{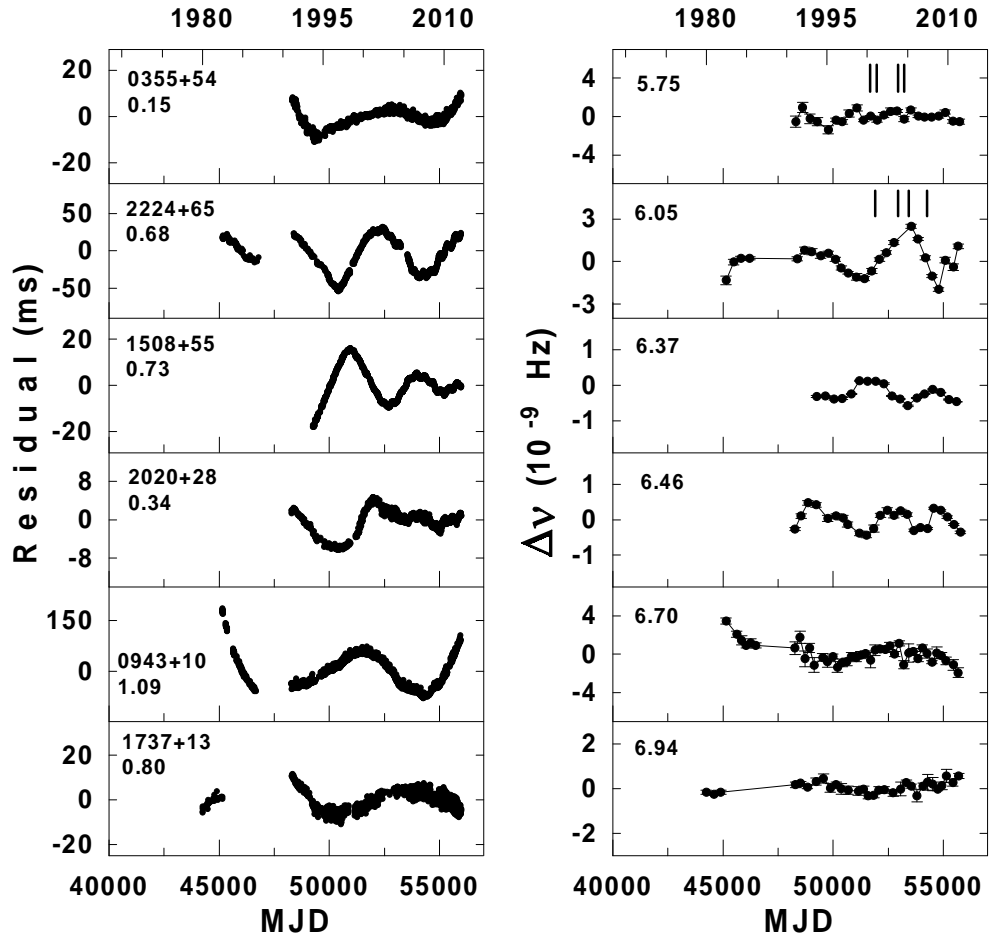}
 \caption{Timing and frequency residuals after the removal of
      a cubic term from the arrival times for the pulsars that show
      cubic components in their timing residuals. The labels are
      the same as in Figure~\ref{resid27}.
     \label{cubic6}}
\end{figure}

\newpage
\clearpage
\begin{figure}
\epsscale{.80}
\plotone{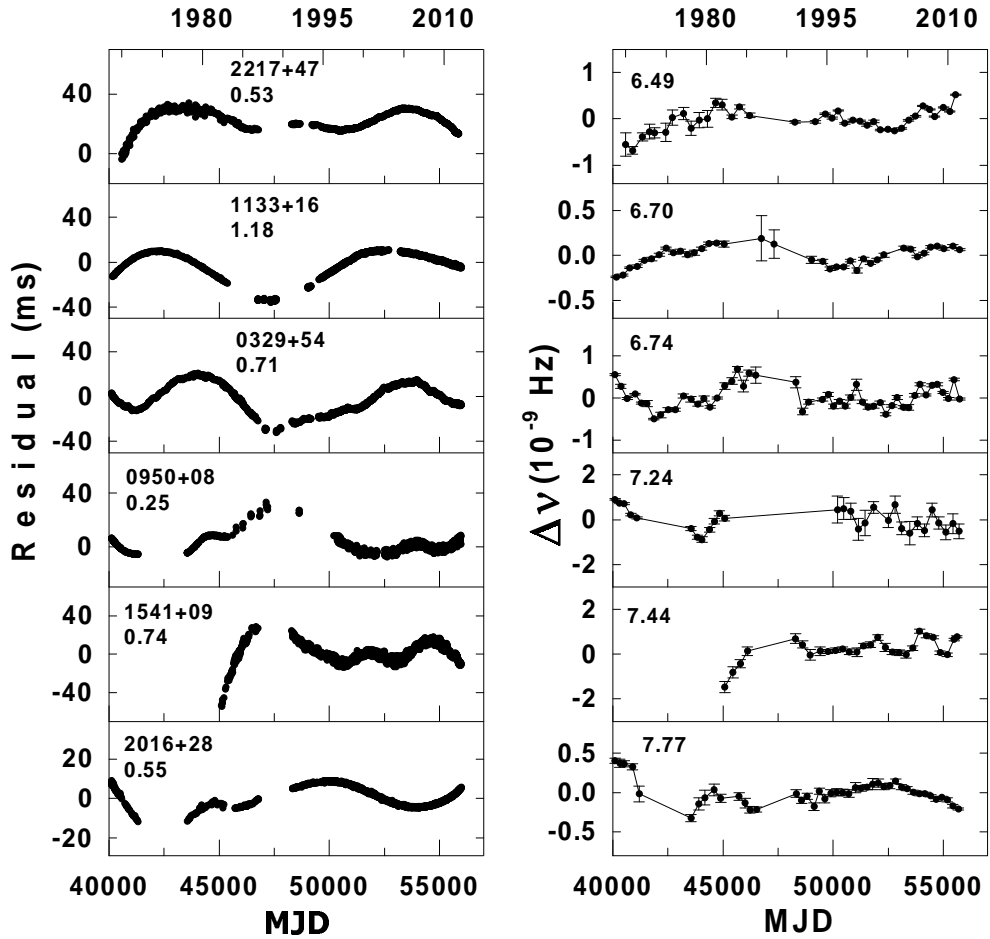}
 \caption{Timing and frequency residuals after the removal of
      a cubic term from the arrival times for the pulsars that show
      cyclical signatures in their timing residuals. The labels are
      the same as in Figure~\ref{resid27}.
     \label{cyclic7}}
\end{figure}

\newpage
\clearpage
\begin{figure}
\epsscale{.80}
\plotone{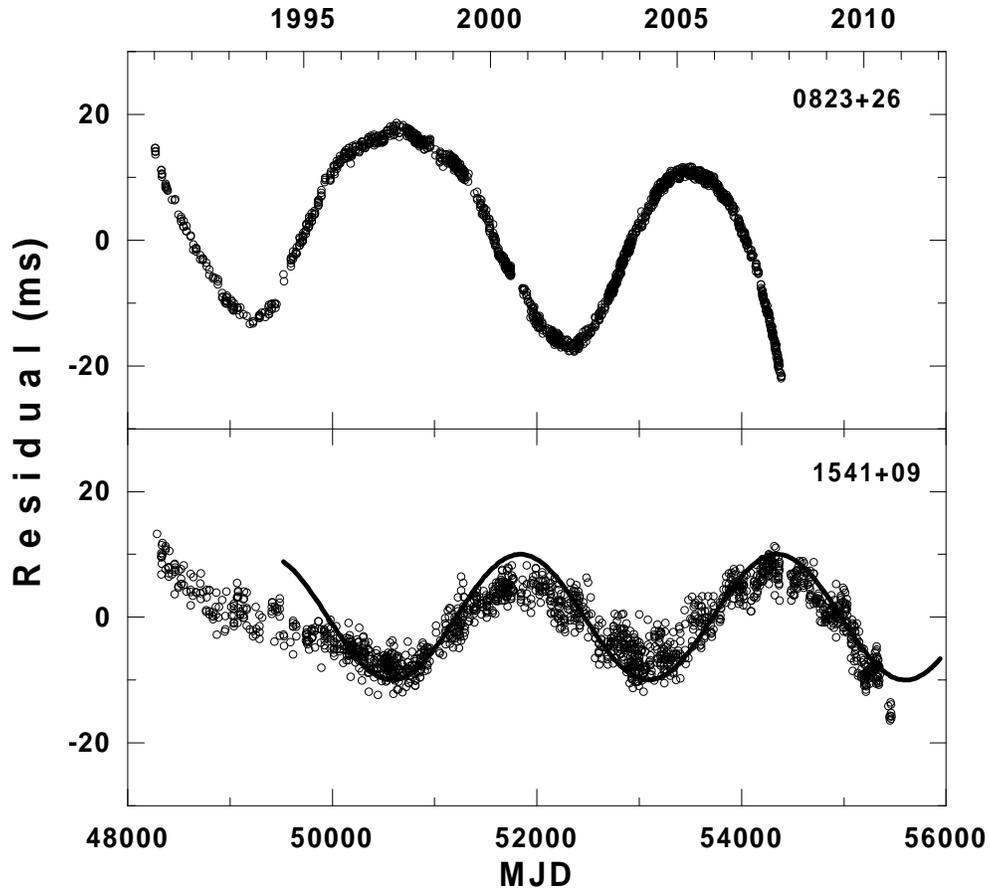}
 \caption{Timing residuals for two pulsars, B0823+26 and B1541+09,
      that exhibit two cycles of clear quasi-sinusoidal structure
      relative to a simple spin-down model. A sinusoidal function
      with a period of $\sim$ 7 yr and an amplitude of 10 ms (marked
      by the bold line) is superimposed on the timing residuals
      of B1541+09. These quasi-periodic structures are not repeated
      in further observations.
     \label{sin1541}}
\end{figure}

\newpage
\clearpage
\begin{figure}
\epsscale{.80}
\plotone{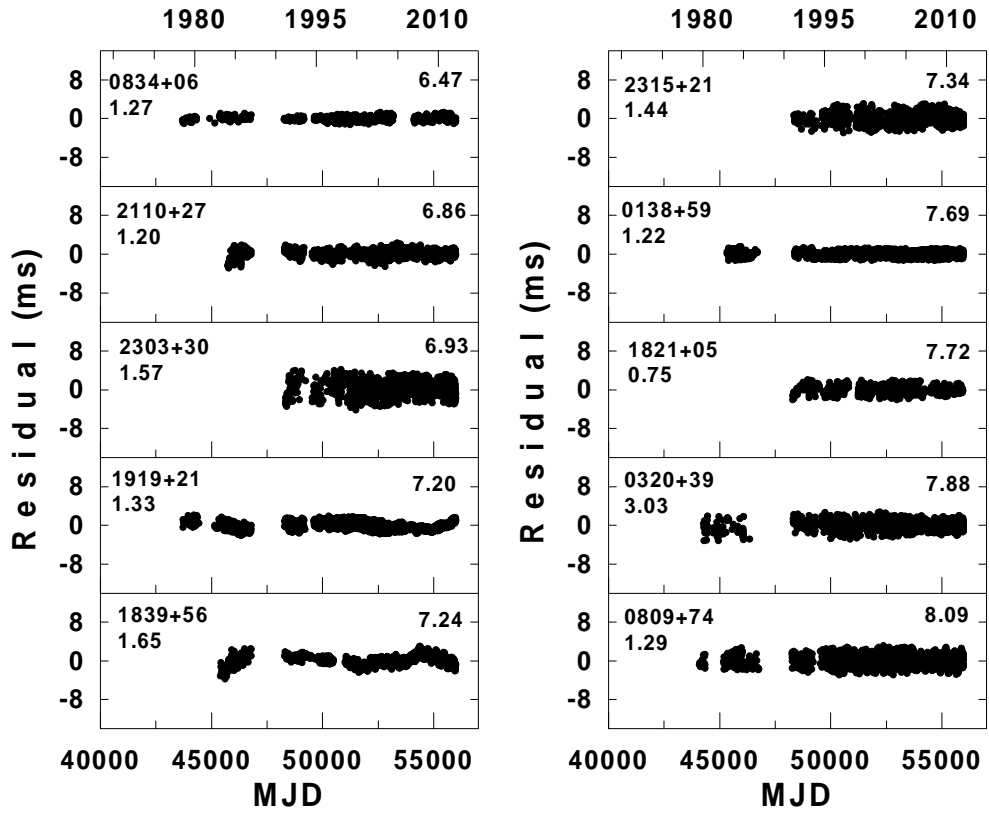}
 \caption{Timing residuals after the removal of a cubic term
      from the arrival times for the pulsars that show noise-like
      variations in their timing residuals. The labels are the same
      as in Figure~\ref{resid27}.
     \label{quiet10}}
\end{figure}

\newpage
\clearpage
\begin{figure}
\epsscale{.70}
\plotone{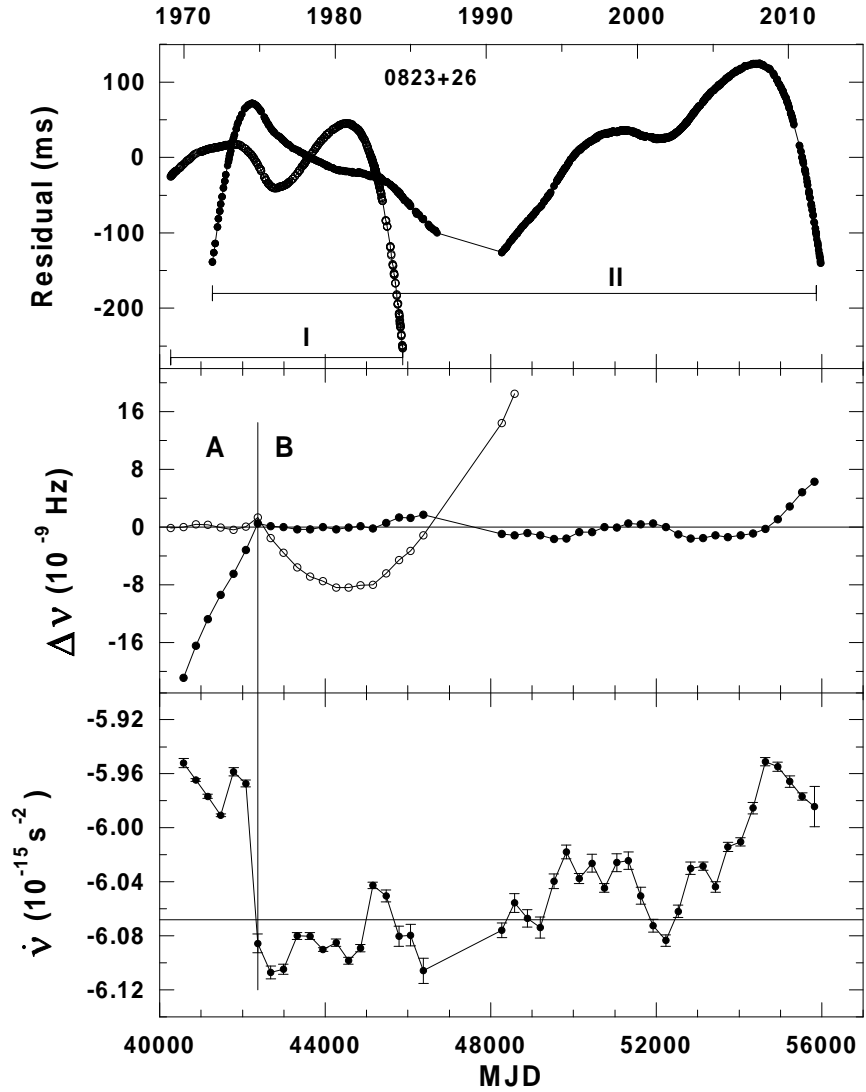}
 \caption{Timing behavior of PSR B0823+26 between 1969 and 2012
     showing a change in the rotation parameters and a change in the sign
     of $\ddot\nu$. Top panel: the residuals relative to two timing models,
     including $\ddot\nu$ with the opposite signs. The two thin
     horizontal lines (I and II) indicate the maximum length of each
     interspace (1969--1984; open circles and 1971--2012; solid circles,
     respectively), in which the data set can be described by each model
     within half a pulse period. Middle panel: the $\Delta\nu$
     residuals relative
     to the two $\nu,\,\dot\nu,\,\ddot\nu$ models: 1969--1974
     (section A before the break; open circles) and 1974--2012
     (section B after the break; solid circles). The time of
     the break in $\Delta\nu$ is marked by the vertical line. 
     Bottom panel: $\dot\nu$ as a function of time. The break in
     $\Delta\nu$ produces a distinct jump in $\dot\nu$.
     \label{sign0823}}
\end{figure}
\newpage
\clearpage
\begin{figure}
\epsscale{.70}
\plotone{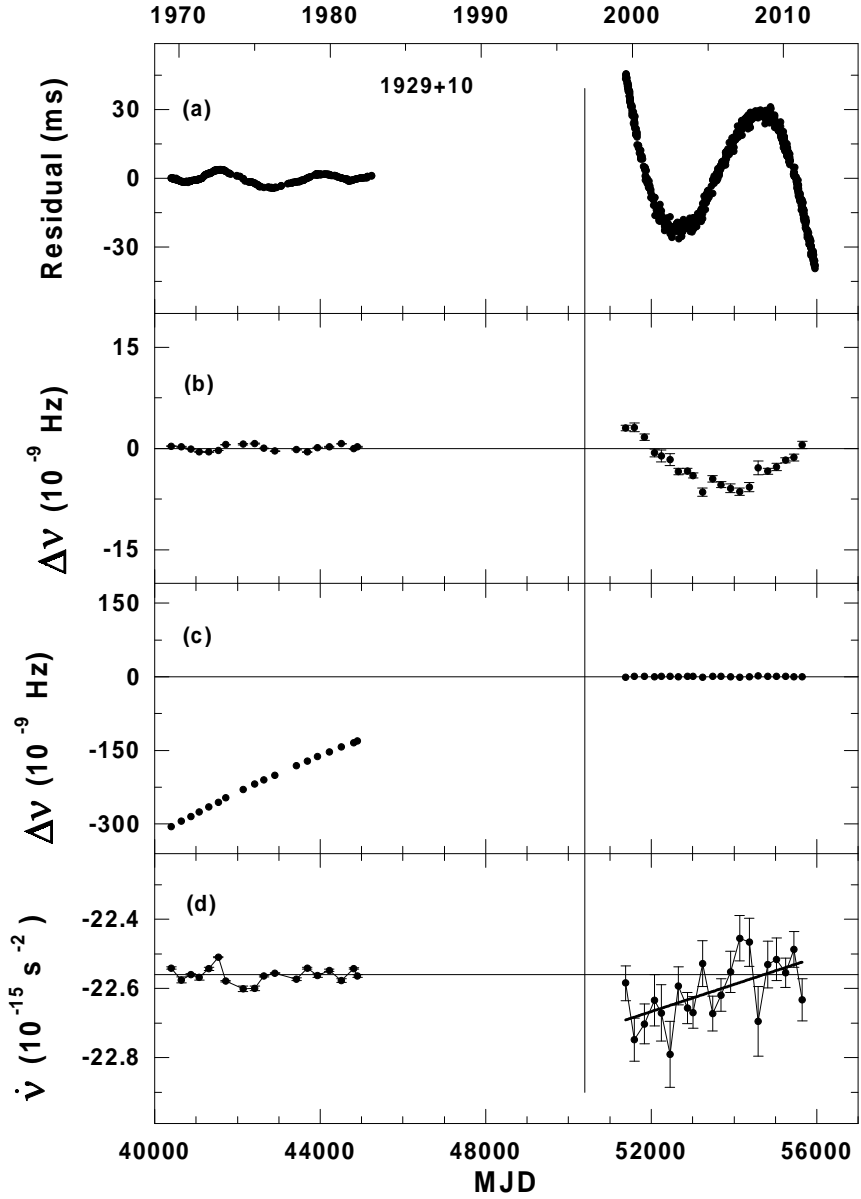}
 \caption{Timing behavior of PSR B1929+10 between 1969 and 2012
     showing a change in the rotation parameters and
     a change in the sign of $\ddot\nu$.
     A 16 yr gap in the data is seen between 1982 and 1999.
     (a) The residuals relative to two simple $\nu,\,\dot\nu$
     models: 1969--1982 (on the left side) and 1999--2012 (on the
     right side). The vertical line marks the approximate time of the
     change of the rotation parameters. (b) The $\Delta\nu$ residuals
     relative to the $\nu,\,\dot\nu,\,\ddot\nu$ timing model
     1969--1982. (c) The $\Delta\nu$ residuals relative to the
     $\nu,\,\dot\nu,\,\ddot\nu$ timing model 1999--2012.
     (d) $\dot\nu$ as a function of time. The best fit 
     linear function is marked by the bold line.
     \label{sign1929}}
\end{figure}
\newpage
\clearpage
\begin{figure}
\epsscale{.80}
\plotone{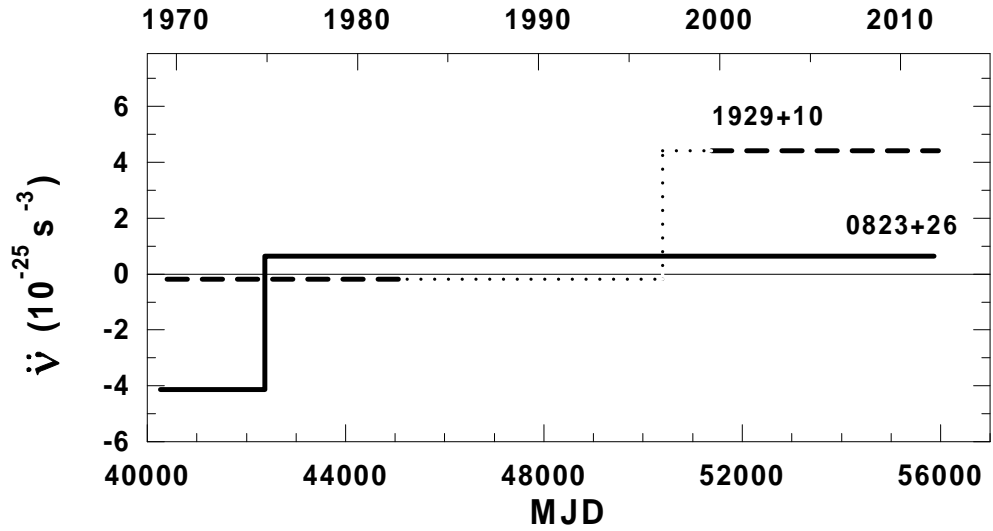}
 \caption{Scheme of the time behavior of $\ddot\nu$ over the interval
     1968--2012 for two pulsars, B0823+26 and B1929+10. The observed change
     in the value and the sign of $\ddot\nu$ is due to a rapid change
     of the pulsar rotation parameters. We see that the time
     behavior of $\ddot\nu$ for PSR B0823+26 is well described by
     a rectangular function (the bold line). The behavior of $\ddot\nu$
     for PSR B1929+10 is plotted by the same function (the dashed line).
     A 16 yr gap in the data between 1982 and 1999 is marked by
     the dotted line. The measured values of $\ddot\nu$ and
     the corresponding MJD ranges are taken from Table~\ref{param27}.
     \label{modulff}}
\end{figure}
\newpage
\clearpage
\begin{figure}
\epsscale{.80}
\plotone{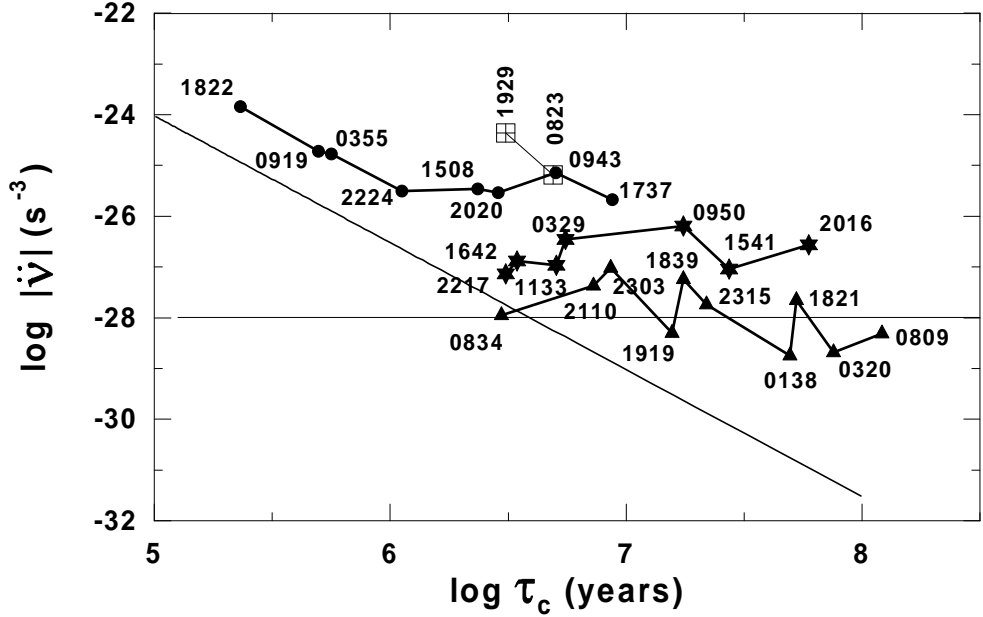}
 \caption{Relation between the measured value of $\ddot\nu$ and
  the characteristic age $\tau_{c}$ for 27
  studied pulsars. Three groups of pulsars form three sequences
  that correspond to three types of signatures of the timing
  residuals. We see that certain signatures of the timing
  residuals are correlated with certain values of $\ddot\nu$ and
  certain age ranges. The pulsars with residuals dominated by
  a cubic trend are marked by filled circles, those pulsars with
  residuals dominated by a quasi-periodic structure are marked by
  star symbols, and those pulsars with residuals dominated by noise-like
  variations are marked by triangle symbols. The two pulsars showing
  a change in the sign of $\ddot\nu$ are marked by squares.
  The thin horizontal line denotes the boundary of accuracy of our
  measurements, equal to $1\times 10^{-28}$ s$^{-3}$.
  The thin sloping line shows an evolutionary trend in accordance
  with a relation $\dot\nu\propto{-{\nu}^{n}}$, where $n=3$.
     \label{zavis27}}
\end{figure}
\newpage
\clearpage
\begin{figure}
\epsscale{.70}
\plotone{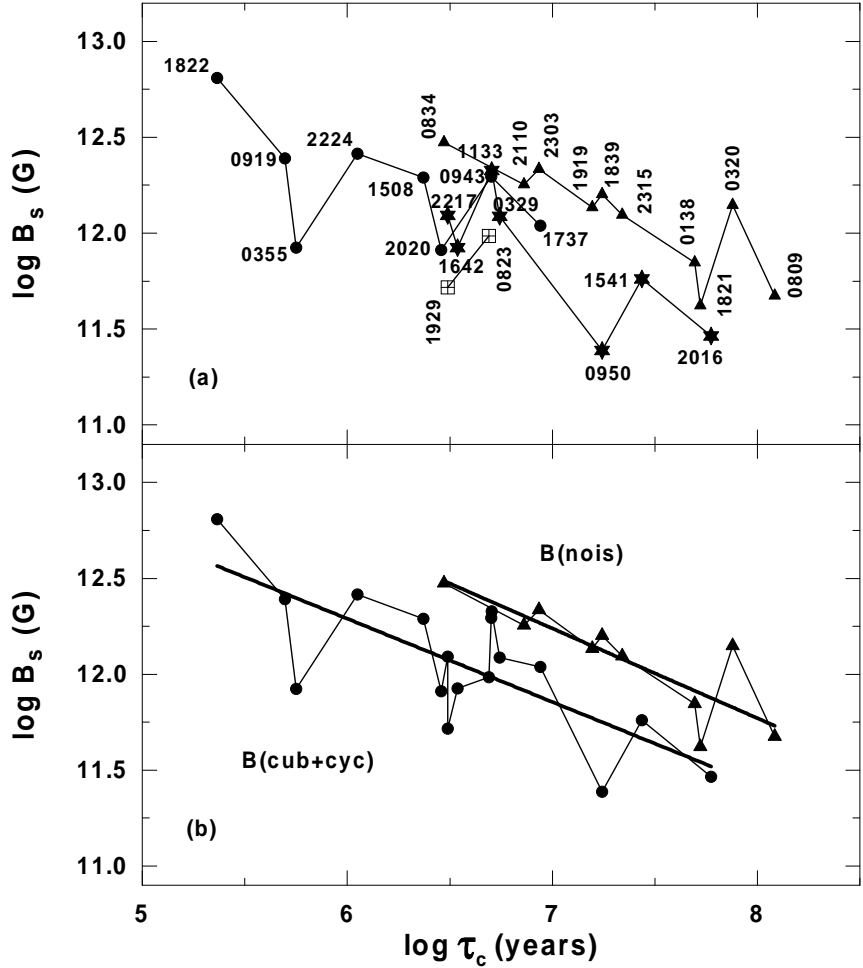}
 \caption{Relation between the surface magnetic field strength
     $B_{s}$ and the characteristic age $\tau_{c}$ for 27 studied
     pulsars. (a) The $B_{s}$ values for the pulsars with cubic
     signatures in their residuals are marked by solid circles
     (the $B_{(cub)}$ values), pulsars with
     quasi-periodic signatures in their residuals are marked by star
     symbols (the $B_{(cyc)}$ values), and  pulsars
     with noise-like signatures in their residuals are marked by
     triangle symbols (the $B_{(nois)}$ values).
     The three plotted sequences indicate that the $B_{s}$
     values are correlated with certain signatures of
     the timing residuals. (b) The $B_{s}$
     points are plotted as in the top panel (a), but the $B_{(cub)}$
     points and the $B_{(cyc)}$ points are united in one sequence
     of the $B_{(cub+cyc)}$ points marked by solid circles.
     The two straight lines fitted to the two sequences of the $B_{s}$
     points show that the $B_{(nois)}$ values are nearly four times
     greater than the $B_{(cub+cyc)}$ values for pulsars
     of the same age.
     \label{zavb0}}
\end{figure}
\newpage
\clearpage
\begin{figure}
\epsscale{.70}
\plotone{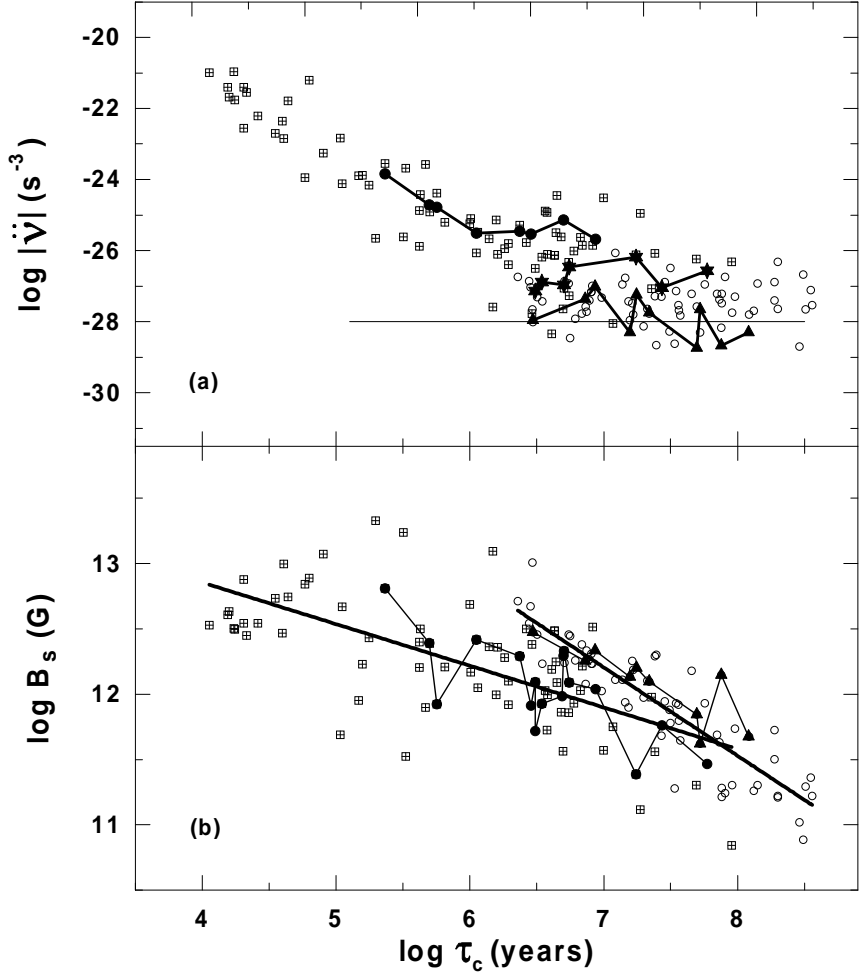}
 \caption{Relation between the $\ddot\nu$ value, the $B_{s}$ value,
    and the characteristic age $\tau_{c}$ for the pulsars from the JBO
    sample. Here, 73 selected pulsars with cubic or quasi-periodic
    components in their residuals are marked by squares
    and 64 pulsars with noise-like components in
    their residuals are marked by open circles.
    (a) Relation between the $\ddot\nu$ value and $\tau_{c}$.
    The three plotted sequences marked by solid lines present
    the PRAO data, as in Figure~\ref{zavis27}. We see that the
    distribution of the JBO points does not contradict the idea of
    three evolutionary stages in pulsar rotation.
    (b) Relation between the $B_{s}$ value and $\tau_{c}$.
    The two sequences marked by solid lines show
    the PRAO data, as in Figure~\ref{zavb0}(b). The two straight
    lines fit to two groups of the JBO points show that
    the magnetic fields are stronger for pulsars exhibiting
    noise-like signatures in their timing residuals.
     \label{zavjbo}}
\end{figure}
\clearpage
\begin{deluxetable}{lcccccc}
\tablecaption{Information on PRAO Pulse Arrival Time Measurements
       \label{inform27}}
\tablewidth{0pt}
\tablehead{
\colhead{PSR J} & \colhead{PSR B} & \colhead{Range} &
\colhead{No of TOAs} & \colhead{Error} &
\colhead{DM} &\colhead{Reference}  \\
\colhead{} & \colhead{} & \colhead{(MJD)} & \colhead{} &
\colhead{(ms)}&
\colhead{(cm$^{-3}$ pc)} & \colhead{}
}
\startdata
 J0141+6009  &B0138+59   &45326-55957  &1541 &0.8  &34.91    &*     \\
 J0323+3944  &B0320+39   &44268-55973  &2660 &0.7  &26.17    &*     \\
 J0332+5434  &B0329+54   &43702-55973  &3037 &0.05 &26.776   &1     \\
 J0358+5413  &B0355+54   &45086-55973  &2632 &0.7  &57.153   &2     \\
 J0814+7429  &B0809+74   &44087-55973  &2777 &0.4  &5.7513   &1     \\
 J0826+2637  &B0823+26   &44807-55973  &2706 &0.2  &19.454   &2     \\
 J0837+0610  &B0834+06   &43701-55973  &2211 &0.1  &12.8579  &1     \\
 J0922+0638  &B0919+06   &45569-55974  &2822 &0.5  &27.271   &2     \\
 J0946+0951  &B0943+10   &45134-55957  &1573 &0.7  &15.4     &2     \\
 J0953+0755  &B0950+08   &43703-55958  &917  &0.2  &2.958    &2     \\
 J1136+1551  &B1133+16   &46750-55974  &2191 &0.2  &4.8471   &1     \\
 J1509+5531  &B1508+55   &49250-55972  &2625 &0.05 &19.613   &2     \\
 J1543+0929  &B1541+09   &45088-55972  &2723 &1.0  &34.99    &1     \\
 J1645$-$0317&B1642$-$03 &45962-55973  &2706 &0.3  &35.737   &*     \\
 J1740+1311  &B1737+13   &48285-55973  &1510 &1.1  &48.673   &2     \\
 J1823+0550  &B1821+05   &48284-55950  &579  &0.9  &66.775   &2     \\
 J1825$-$0935&B1822$-$09 &48333-55973  &1713 &1.0  &19.383   &*     \\
 J1840+5640  &B1839+56   &45398-55973  &1743 &0.2  &26.764   &*     \\
 J1921+2153  &B1919+21   &43715-55973  &2187 &0.2  &12.4309  &1     \\
 J1932+1059  &B1929+10   &51385-55957  &946  &0.4  &3.180    &2     \\
 J2018+2839  &B2016+28   &45724-55973  &2327 &0.2  &14.19    &*     \\
 J2022+2854  &B2020+28   &48270-55973  &2061 &0.4  &24.66    &*     \\
 J2113+2754  &B2110+27   &45724-55973  &2447 &0.5  &25.113   &2     \\
 J2219+4754  &B2217+47   &45073-55973  &1491 &0.06 &43.519   &2     \\
 J2225+6535  &B2224+65   &45144-55957  &1281 &0.8  &36.226   &*     \\
 J2305+3100  &B2303+30   &48329-55973  &955  &1.2  &49.544   &2     \\
 J2317+2149  &B2315+21   &48330-55971  &1309 &1.1  &20.854   &*     \\
\enddata
\tablecomments{In the column order, the table shows the pulsar
  name (J2000 and B1950),
  the MJD range of PRAO timing measurements, the number of times
  of arrival (TOA) in each range, the mean measurement error for
  observing session lasting from 3 to 11 minutes for
  different pulsars, and the dispersion measure parameter DM.
  The references in the last column indicate the DM value that provided
  the best adjustment between the two sets of the timing residuals
  obtained for each pulsar at the two observing frequencies of
  102.7 MHz (until 1998 May) and 111.3 MHz (since 1998 November) in
  the PRAO observations. }
\tablerefs{(1) \citet{tay93}; (2) \citet{hob04}; (*) this work. }
\end{deluxetable}
\clearpage
\begin{deluxetable}{llllcccc}
\tablecaption{Observed Parameters for 27 Pulsars
       \label{param27}}
\tablewidth{0pt}
\tablehead{
\colhead{PSR B} & \colhead{$\nu$} &
\colhead{$\dot\nu$} & \colhead{$\ddot\nu$} & \colhead{Epoch} &
\colhead{Range} & \colhead{$\sigma_{2}$} & \colhead{$\sigma_{3}$} \\
\colhead{} & \colhead{(s$^{-1}$)} &
\colhead{($10^{-15}$ s$^{-2}$)} &
\colhead{($10^{-25}$ s$^{-3}$)} & \colhead{(MJD)} &
\colhead{(MJD)} & \colhead{(ms)}  & \colhead{(ms)}
}
\startdata
   B0138+59     &0.8176959968288(5)  &-0.261461(3) &0.00018(6)
            &45326.7082 &45326-55957 &0.6  &0.5        \\
   B0320+39     &0.329807506325(1)   &-0.069141(3) &0.00021(5)
        &44268.6762 &44268-55973 &1.0  &1.0         \\
   B0329+54     &1.399543746342(8)   &-4.01374(3)  &0.0343(4)
        &40105.2304 &40105-55973 &24     &11        \\
   B0355+54     &6.39460122291(2)    &-179.7892(2) &1.671(4)
            &48298.6660 &42298-55973 & ...   &2.7         \\
   B0809+74     &0.773849236501(1)   &-0.100671(4) &0.00049(8)
        &44087.3634 &44087-55973 &1.2  &1.2         \\
   B0823+26     &1.88444719685(3)    &-5.929(1)    &-4.14(8)
        &40264.4381 &40264-42364 &4.9   &1.4        \\
   B0823+26     &1.88444611497(2)    &-6.0976(1)   &0.646(2)
        &42364.6405 &42364-55973 &  ...   &40.2       \\
   B0834+06     &0.7850739585879(2)  &-4.190673(1) &0.00113(3)
        &43701.4411 &43701-55973 &0.3    &0.2       \\
   B0919+06     &2.32226169326(9)    &-74.0527(5)  &1.91(2)
            &44210.5817 &44210-55139 & ...     &57        \\
   B0919+06     &2.32219477708(9)    &-74.037(5)   &18(1)
            &55140.1604 &54140-55974 &1.9    &2.2       \\
   B0943+10     &0.91099181885(4)    &-2.8750(2)   &-0.719(4)
        &45134.5681 &45134-55957 & ...      &48       \\
   B0950+08     &3.951553316192(9)   &-3.58546(3)  &-0.0644(6)
        &40104.8560 &40104-55958 &20.7   &5.1       \\
   B1133+16     &0.841813872293(3)   &-2.64632(1)  &0.0109(2)
        &40154.6631 &40154-55974 &16      &8.6      \\
   B1508+55     &1.351932973645(13)  &-9.13014(9)  &-0.347(4)
        &49250.5223 &49250-55972 &18      &6        \\
   B1541+09     &1.33609766492(2)    &-0.77321(6)  &0.009(2)
        &45087.9517 &45088-55972 &8.9     &8.8      \\
   B1642$-$03   &2.57938868630(3)    &-11.8466(2)  &0.013(3)
        &40414.1297 &40414-55973 &23     &23        \\
   B1737+13     &1.245252847252(4)   &-2.25742(2)  &0.2121(3)
        &44240.7561 &44240-55973 &52     &4           \\
   B1821+05     &1.328186108116(4)   &-0.40025(2)  &0.0022(6)
        &48284.3024 &48284-55950 &0.9    &0.9       \\
   B1822$-$09   &1.30038133415(4)    &-88.859(2)   &14.4(1)
            &54116.3366 &54116-55973 &14.8   &4.5       \\
   B1839+56     &0.6050113973913(7)  &-0.547042(3) &-0.0057(2)
        &45398.2170 &45398-55973 &2.4    &1.0       \\
   B1919+21     &0.747774504487(1)   &-0.753867(2) &-0.00050(5)
        &43715.8588 &43715-55973 &0.9    &0.8      \\
   B1929+10     &4.41467924848(3)    &-22.5547(4)  &-0.18(2)
        &40401.2136 &40401-45236 &2.1    &2.0       \\
   B1929+10     &4.41465783839(3)    &-22.6674(3)  &4.42(1)
        &51385.8682 &51385-55957 &20.5   &2.3       \\
   B2016+28     &1.792264372162(6)   &-0.47720(2)  &0.0269(2)
        &40105.0491 &40105-55973 &12.0   &4.3       \\
   B2020+28     &2.912039586312(7)   &-16.06542(4) &0.2926(8)
        &48270.4234 &48270-55973 &9.6     &1.8      \\
   B2110+27     &0.831358116226(1)   &-1.813019(3) &0.00429(7)
        &45724.4286 &45724-55973 &1.3     &0.7      \\
   B2217+47     &1.857122692273(8)   &-9.53658(3)  &-0.0073(3)
        &40585.0654 &40585-55854 &7.1     &6.1      \\
   B2224+65     &1.46512685285(5)    &-20.7457(2)  &0.313(4)
        &45144.0676 &45144-55957 &57      &24       \\
   B2303+30     &0.634563570173(2)   &-1.16455(2)  &-0.0096(5)
        &48329.3741 &48329-55973 &2.3     &1.7      \\
   B2315+21     &0.692207715845(1)   &-0.501662(8) &-0.0018(3)
        &48330.3796 &48330-55971 &1.1     &1.1      \\
\enddata
\tablecomments{In the column order,
  the table shows the pulsar's B1950 name, the pulsar's rotation
  parameters $\nu$, $\dot\nu$, and $\ddot\nu$, the epoch of the $\nu$
  measurement, the MJD range, the rms value remaining after a second-order
  fit $\sigma_{2}$, and the rms value remaining after a third-order
  fit $\sigma_{3}$. Numbers in parentheses are 1-$\sigma$ uncertainties
  in the last digit quoted. The parameters for B0823+26, B0919+06,
  and B1929+10 are shown for two timing models at two
  different intervals.}
\end{deluxetable}

\clearpage
\begin{deluxetable}{lcccc}
\tablecaption{Parameters for Eight Glitches in PSR B1822$-$09
              from the PRAO Measurements
        \label{glit1822}}
\tablewidth{0pt}
\tablehead{
\colhead{No.} & \colhead{Description} & \colhead{Epoch} &
\colhead{${\Delta\nu}/{\nu}$} & \colhead{${\Delta\dot\nu}/{\dot\nu}$} \\
\colhead{}    & \colhead{of the Glitch} & \colhead{(MJD)} &
\colhead{$(10^{-9})$} & \colhead{$(10^{-3})$}
}
\startdata
 1   &1994      &49615(8) &0.2(1)   &-0.6(2)   \\
\tableline
 2   &1  slow   &49857    &12.8(2)  & -7.0(2)   \\
 3   &1a slow   &50253    &4.3(2)   & -4.8(3)   \\
 4   &2  slow   &51060    &28.7(6)  & -24.2(4)  \\
 5   &3  slow   &51879    &32.0(9)  & -16.7(8)  \\
 6   &4  slow   &52700    &2.5(3)   & -2.9(3)   \\
\tableline
 7   &2006      &53745(2)   &6.7(4)  &-0.1(6)  \\
 8   &2007      &54115.0(2) &121(1)  &-0.2(4)  \\
\enddata
\tablecomments{The table lists the current glitch
  number, the year for the normal glitch or the number of
  the slow glitch, the epoch of the normal or slow glitch,
  and the glitch parameters. Uncertainties in the parameters
  are in parentheses and refer to the last digit quoted.}
\end{deluxetable}

\clearpage
\begin{deluxetable}{lccc}
\tablecaption{DM Measurements for PSR B2224+65
       \label{DM}}
\tablewidth{0pt}
\tablehead{
\colhead{DM} & \colhead{Epoch} & \colhead{Frequencies} &
\colhead{References} \\
\colhead{(pc cm$^{-3}$)} & \colhead{(MJD)} & \colhead{(MHz)}&
\colhead{}
}
\startdata
 36.16(5)   &48382  &400, 1640             & 1         \\
 36.079(9)  &49303  &408, 610, 910, 1630   & 2         \\
 36.226(5)  &51100  &102.746, 111.646      & this work \\
 36.42(2)   &52400  &840, 1380             & 3         \\
\enddata
\tablerefs{(1) \citet{arz94}; (2) \citet{hob04}; (3) \citet{jan06}.}
\end{deluxetable}

\clearpage
\begin{deluxetable}{lc}
\tablecaption{Glitch Parameters for PSR B2217+47
        \label{glt2217}}
\tablewidth{0pt}
\tablehead{
\colhead{Glitch Parameters} & \colhead{Values}
}
\startdata
 Pre-glitch parameters            &                   \\
 MJD range                        & 55600--55854      \\
 $\nu$ (Hz)                       & 1.85711010737(1)  \\
 $\dot\nu$ ($10^{-15}$ s$^{-2}$)  & -9.5599(7)        \\
 Epoch (MJD)                      & 55858.0           \\
 rms timing residual ($\mu$s)     & 60                \\
\tableline
 Post-glitch parameters           &                   \\
 MJD range                        & 55866--55973      \\
 $\nu$ (Hz)                       & 1.85711010972(3)  \\
 $\dot\nu$ ($10^{-15}$ s$^{-2}$)  & -9.536(5)         \\
 Epoch (MJD)                      & 55858.0           \\
 rms timing residual ($\mu$s)     & 60                \\
\tableline
 Glitch parameters                      &             \\
 ${\Delta\nu}/{\nu}\,(10^{-9})$         & 1.27(4)     \\
 ${\Delta\dot\nu}/{\dot\nu}\,(10^{-3})$ & -2.5(5)     \\
 Epoch (MJD)                            & 55858(1)    \\
\enddata
\tablecomments{
  The rotation parameters are calculated at the glitch epoch
  MJD 55858.0.}

\end{deluxetable}

\clearpage
\begin{deluxetable}{lcccccc}
\tablecaption{Derived Parameters for 27 Pulsars
       \label{deriv27}}
\tablewidth{0pt}
\tablehead{
\colhead{PSR B} & \colhead{${\tau_{c}}$} &
\colhead{$\log({\tau_{c}})$} & \colhead{$\log({\ddot\nu}_{obs})$} &
\colhead{$n$} &
\colhead{$\log({{\ddot\nu}_{exp}})$} & \colhead{$\log$($B_{s}$)}  \\
\colhead{} & \colhead{(Myr)} & \colhead{(yr)} &
\colhead{($10^{-25}$ s$^{-3}$)} &
\colhead{} & \colhead{($10^{-25}$ s$^{-3}$)} & \colhead{(G)}
}
\startdata
B1822$-$09 &0.23  &5.36  &-23.84  &237   &-25.74 &12.81 \\
B0919+06   &0.50  &5.70  &-24.70  &80    &-26.15 &12.39 \\
B0919+06   &0.50  &5.70  &-23.74  &762   &-26.15 &12.39 \\
B0355+54   &0.56  &5.75  &-24.78  &33    &-25.82 &11.92 \\
B2224+65   &1.10  &6.05  &-25.50  &106   &-27.06 &12.42 \\
B1508+55   &2.30  &6.37  &-25.46  &-562  &-27.73 &12.29 \\
B2020+28   &2.87  &6.46  &-25.53  &330   &-27.58 &11.91 \\
B0834+06   &2.95  &6.47  &-27.96  &4     &-28.17 &12.47 \\
B2217+47   &3.08  &6.49  &-27.14  &-14   &-27.83 &12.09 \\
B1929+10   &3.10  &6.49  &-25.74  &-156  &-27.46 &11.72 \\
B1929+10   &3.08  &6.49  &-24.36  &3797  &-27.46 &11.72 \\
B1642$-$03 &3.45  &6.54  &-26.89  &23    &-27.79 &11.92 \\
B0823+26   &5.04  &6.70  &-24.38  &-22188&-28.25 &11.98 \\
B0823+26   &4.90  &6.69  &-25.19  &3274  &-28.23 &11.98 \\
B0943+10   &5.02  &6.70  &-25.14  &-7924 &-28.56 &12.30 \\
B1133+16   &5.04  &6.70  &-26.96  &131   &-28.60 &12.33 \\
B0329+54   &5.52  &6.74  &-26.46  &297   &-28.46 &12.10 \\
B2110+27   &7.26  &6.86  &-27.37  &108   &-28.93 &12.26 \\
B2303+30   &8.63  &6.93  &-27.01  &-449  &-29.19 &12.34 \\
B1737+13   &8.74  &6.94  &-25.67  &5181  &-28.91 &12.04 \\
B1919+21   &15.71 &7.20  &-28.30  &-65   &-29.64 &12.13 \\
B0950+08   &17.46 &7.24  &-26.19  &-1979 &-29.01 &11.39 \\
B1839+56   &17.52 &7.24  &-27.24  &-1152 &-29.83 &12.20 \\
B2315+21   &21.86 &7.34  &-27.74  &-496  &-29.96 &12.10 \\
B1541+09   &27.37 &7.44  &-27.05  &2012  &-29.87 &11.76 \\
B0138+59   &49.55 &7.69  &-28.74  &215   &-30.60 &11.84 \\
B1821+05   &52.57 &7.72  &-27.66  &1825  &-30.44 &11.62 \\
B2016+28   &59.50 &7.77  &-26.57  &21186 &-30.42 &11.46 \\
B0320+39   &75.57 &7.88  &-28.68  &1455  &-31.36 &12.15 \\
B0809+74   &121.8 &8.09  &-28.31  &3717  &-31.41 &11.67 \\
\enddata
\tablecomments{The table lists the pulsars in order of increasing 
age. The columns list the pulsar B1950 name, the characteristic age
$\tau_{c}={\nu}/{2\dot{\nu}}$ and its logarithmic value,
the logarithmic values of the observed value ${\ddot\nu}_{obs}$,
the observed braking index
$n={\nu{{\ddot\nu}_{obs}}}/{{\dot\nu}^{2}}$, the expected value
${{\ddot\nu}_{exp}}={3{\dot\nu}^{2}}/{\nu}$ on a logarithmic scale,
and the logarithmic value of surface magnetic field strength
$B_{s}=3.2\times{10^{19}}({P}{\dot{P}})^{1/2}$ G.
The parameters for B0823+26, B0919+06, and B1929+10 are shown
for two different models according to Table~\ref{param27}.}

\end{deluxetable}
\end{document}